\documentclass[epj,nopacs]{svjour}

\usepackage[dvipdfmx]{graphicx}
\usepackage{subfigure}
\usepackage{amsmath,amssymb}
\usepackage{bm}
\usepackage{scalefnt}
\usepackage{color}
\usepackage{slashed}
\usepackage[
 colorlinks=true,
 linkcolor=red,
 citecolor=blue,
]{hyperref}
\usepackage{comment}
\usepackage{latexsym}
\usepackage{relsize}
\def\Babar{{\mbox{\slshape B\kern-0.1em{\smaller A}\kern-0.1em B\kern-0.1em{\smaller A\kern-0.2em R}}}}

\newcommand{\Dgen}{{D^{(*)}}}

\newcommand{\GeV}{\text{GeV}}

\renewcommand{\arraystretch}{1.2}

\begin{document}
\title{\boldmath Testing Leptoquark/EFT in $\bar B \to \Dgen l\bar\nu$ at the LHC}
\titlerunning{Testing Leptoquark/EFT in $\bar B \to \Dgen l\bar\nu$ at the LHC}
\authorrunning{S.~Iguro et al.}

\author{Syuhei Iguro\inst{1} \and Michihisa Takeuchi\inst{2} \and Ryoutaro Watanabe\inst{3,4} }                     
\institute{Department of Physics, Nagoya University, Nagoya 464-8602, Japan 
\and Kobayashi-Maskawa Institute for the Origin of Particles and the Universe, Nagoya University, Nagoya 464-8602, Japan
\and INFN, Sezione di Roma Tre, Via della Vasca Navale 84, 00146 Rome, Italy
\and INFN, Sezione di Pisa, Largo B. Pontecorvo 3, 56127 Pisa, Italy (as a visitor)}
\date{Update: August 20, 2022~~(Received: April 12, 2021)}

\abstract{
We investigate the current LHC bounds on New Physics (NP) that contributes to $\bar B \to \Dgen l\bar\nu$ for $l = (e,\mu,\tau)$ 
by considering both leptoquark (LQ) models and an effective-field-theory (EFT) Hamiltonian. 
Experimental analyses from $l+\text{missing}$ searches with high $p_T$ are applied to evaluate the NP constraints with respect to the Wilson coefficients. 
A novel point of this work is to show difference between LQs and EFT for the applicable LHC bound. 
In particular, we find that the EFT description is {\it not} valid to search for LQs with the mass less than $\lesssim 10\,\text{TeV}$ at the LHC 
and leads to overestimated bounds. 
We also discuss future prospects of high luminosity LHC searches including the charge asymmetry of background and signal events. 
Finally, a combined summary for the flavor and LHC bounds is given, and then we see that in several NP scenarios the LHC constraints are comparable with the flavor ones. 
\\
%
} 

\maketitle
\section{\boldmath Introduction}
\label{sec:intro}

The flavor studies of the semi-leptonic $B$ decay processes have been developed these years to test the electroweak sector of the Standard Model (SM) and also to look for evidence of New Physics (NP) effects. 
In particular, the charged current is described as 
\begin{align}
 \mathcal H_\text{eff} = 2\sqrt2 G_F V_{qb} C_\text{model} \, (\bar q \Gamma^\mu b) (\bar l \Gamma_\mu \nu) \,,
\end{align}
with $C_\text{SM} = 1$ and $\Gamma^\mu = \gamma^\mu P_L$ in the SM, where $G_F$, $V_{qb}$, and $P_{L/R}$ denote the Fermi coupling constant, the Cabibbo-Kobayashi-Maskawa~\cite{Cabibbo:1963yz,Kobayashi:1973fv} (CKM) matrix element, and chiral projection operators $(1\mp\gamma_5)/2$, respectively. 
Throughout this paper, $l$ indicates all the charged leptons ($l=e,\mu,\tau$), whereas $\ell$ represents the light charged leptons ($\ell = e,\mu$). 
Then, the NP effect in the Wilson coefficient (WC) $C_\text{NP}$, with an arbitrary Lorenz structure $\Gamma$, can be analyzed from the $B$ decay observables.

The exclusive processes of $\bar B \to \Dgen \ell \bar\nu$ and $\bar B \to \pi \ell \bar\nu$ are used to determine $|V_{cb}|$ and $|V_{ub}|$, respectively. 
In Ref.~\cite{Iguro:2020cpg}, the authors extended the fit analysis for the $|V_{cb}|$ determination to include NP effects on the decay distributions with the help of the updated treatment of the form factors~\cite{Bordone:2019vic}. 
Then, it turns out that non-zero NP contributions with the size of $C_\text{NP} \sim \mathcal O (\%)$ are possible in the $b \to c\ell\nu$ current, as will be shown later.
A similar study for $b \to u\ell\nu$ would be available in future.

On the other hand, the tau modes $\bar B \to \Dgen \tau \bar\nu$ are of particular interest 
because of the excesses in the experimental measurements of $R_{\Dgen} = \mathcal B (\bar B \to \Dgen \tau\bar\nu) / \mathcal B (\bar B \to \Dgen \ell\bar\nu)$ compared with the SM predictions. 
The current measurements are summarized as $R_{\scalebox{0.7}{$D$}}^\text{\scalebox{0.7}{exp:WA}} = 0.340(27)(13) $ and $R_{\scalebox{0.7}{$D^*$}}^\text{\scalebox{0.7}{exp:WA}} = 0.295(11)(8)$ 
while several SM predictions are reported, {\it e.g.}, $R_{\scalebox{0.7}{$D$}}^\text{\scalebox{0.7}{SM}} = \{0.299(4),\, 0.297(6),\, 0.289(4)\}$ and $R_{\scalebox{0.7}{$D^*$}}^\text{\scalebox{0.7}{SM}} = \{0.258(5),\, 0.245(4),\, 0.248(1)\}$, 
where the first ones are from HFLAV~\cite{Amhis:2019ckw} and the latter two are from Ref.~\cite{Iguro:2020cpg} based on two different sets for the form factors. 
Then, many studies point out that the excesses can be explained by several types of NP currents in $b \to c\tau\nu$ with $C_\text{NP} \sim \mathcal O (10\%)$.  
There also exists the NP study for $b \to u\tau\nu$ as in Ref.~\cite{Tanaka:2016ijq}, and then the typical size of the constraint is $\mathcal O (10\%)$ as well.

In light of the above situation, it is natural to ask if we can access such NP effects of $1\,\text{--}\,10\%$ order at the large hadron collider (LHC) searches. 
In Ref.~\cite{Greljo:2018tzh}, $\tau + \text{missing}$ searches by ATLAS~\cite{Aaboud:2018vgh} and CMS~\cite{Sirunyan:2018lbg} have been applied to put the LHC bound on $C_\text{NP}$ in the $b \to c\tau\nu$ current. 
Then, they found that the current LHC data constrains NP scenarios addressing the $R_{\Dgen}$ excesses. 
See also Refs.\cite{Dumont:2016xpj,Altmannshofer:2017poe,Iguro:2017ysu,Abdullah:2018ets,Iguro:2018fni,Baker:2019sli,Marzocca:2020ueu} for the other studies. 
One can think that the $\ell + \text{missing}$ search by ATLAS with $139~\text{fb}^{-1}$~\cite{Aad:2019wvl} can probe the LHC bound in $b \to q \ell\nu$ with $q = c, u$ as well.

In this work, we will obtain the LHC bounds for all the possible types of the NP currents in $b \to q l\nu$. 
{\it A novel point} of this work is, however, not only for such a comprehensive analysis, but rather for pointing out difference between the Effective-Field-Theory (EFT) and actual NP models, as bellow.

A common outlook on these LHC analyses is that NP constraints are obtained only from a high $p_T$ tail of a SM background (BG) due to a $W$ resonance. 
To be precise, the transverse mass with $m_T \sim 2\,\text{--}\,3\,\text{TeV}$ is sensitive to the NP contributions. 
In this case, the EFT description is not always appropriate for actual NP models, whose effect is usually encoded in $C_\text{NP}$, as will be shown in this work. 
We will clarify this point and show that the LHC bound depends on the NP particle mass in the WC.

For this purpose, we focus on NP models with non-resonant $m_T$ distribution, namely, leptoquark (LQ) models.
Eventually, we will show that the EFT limit is {\it not} applicable for the LQ particle mass less than $\mathcal O (10) \,\text{TeV}$ due to angular and energy dependence of the charged lepton $l^\pm$, which might not much be paid attention so far.

Our paper is organized as follows. 
In Sec.~\ref{sec:EFT} and Sec.~\ref{sec:LQ}, we define the model independent NP interactions in $b \to q l\nu$ and the corresponding LQ models, along with the summary of the current flavor bounds. 
In Sec.~\ref{sec:collider}, we show the $m_T$ distribution of the cross section, and see how the EFT and the LQ model differ at the high $p_T$ tail. 
Then we present our analysis for the LHC bound on the NP contribution. 
In Sec.~\ref{sec:discussion}, we compare the flavor and LHC bounds, and indicate significance of non-EFT constraints. 
Finally, our summary and discussion are given in Sec.~\ref{sec:conclusion}.

\section{\boldmath Effective Field Theory and flavor bound}
\label{sec:EFT}

In this work, we start with the weak EFT basis Hamiltonian for the semi-leptonic process $b \to ql\nu$: 
\begin{align}
 \label{Eq:effH}
 {\mathcal H}_{\rm{eff}}=2 \sqrt 2 G_FV_{qb} \Bigl[ 
 & (1 + C_{V_1}^{ql}) (\overline{q} \gamma^\mu P_Lb)(\overline{l} \gamma_\mu P_L \nu_{l}) \notag \\
 & +C_{V_2}^{ql} (\overline{q} \gamma^\mu P_Rb)(\overline{l} \gamma_\mu P_L \nu_{l}) \notag \\[0.2em]
 & +C_{S_1}^{ql} (\overline{q} P_Rb)(\overline{l} P_L \nu_{l}) \notag \\[0.2em]
 & +C_{S_2}^{ql} (\overline{q} P_Lb)(\overline{l} P_L \nu_{l})  \\
 & +C_{T}^{ql} (\overline{q}  \sigma^{\mu\nu}P_Lb)(\overline{l} \sigma_{\mu\nu} P_L \nu_{l}) \Bigl] + \,\text{h.c.}. \notag
\end{align} 
Then, the NP contributions are involved in the WCs $C_X^{ql}$ with the SM normalization $2 \sqrt 2 G_FV_{qb}$. 
In this work, we take $| V_{cb} | = 0.0410(14)$ and $| V_{ub} | = 0.00382(24)$ from PDG2020~\cite{Zyla:2020zbs}.  
We do not consider the right-handed neutrinos.

The $B$ meson decays are described with respect to the WCs at a low energy scale $\mu = m_b$ while a NP model is set at a scale $\mu = \Lambda$. 
At flavor physics, the EFT limit $q^2 \ll \Lambda^2$ is a good approximation for $\Lambda \gtrsim \mathcal O(100)\,\text{GeV}$. 
In this case, the corresponding WC is given as 
\begin{align}
 \label{eq:WCEFT}
 2 \sqrt 2 G_FV_{qb} C_X (\Lambda) = N_X {h_1 h_2 \over M_\text{NP}^2} \,,
\end{align}
with a mass of a NP particle $M_\text{NP}$ and its couplings to the SM fermions $h_{1,2}$, and one may typically take $\Lambda = M_\text{NP}$. 
The numerical factor $N_X$ depends on the Lorenz structure of the EFT operator. 
Then, the WCs at these two scales, $C_X (m_b)$ and $C_X (\Lambda)$, are connected by renormalization-group equation (RGE). 
In this work we follow Ref.~\cite{Iguro:2018vqb} for the formula.

Existing flavor bounds on $C_X (m_b)$ are summarized as below. 
\begin{itemize}
 \item 
 $b \to c \ell\nu$: 
 the comprehensive fit analysis~\cite{Iguro:2020cpg} of the semi-leptonic processes $\bar B \to \Dgen \ell\bar\nu$ points to non-zero preferred values of the WCs for $V_2$ and $T$. 
 The fit along with the form factors leads to $C_{V_2}^{c\ell} (m_b) = +0.02(1)$, $+0.05(1)$ and $C_{T}^{c\ell} (m_b) = +0.02(1)$ depending on the form factor description. 
 See Ref.~\cite{Iguro:2020cpg} for detail. \\[-0.7em]
 \item 
 $b \to c \tau\nu$:
 the excesses of the $R_{\Dgen}$ measurements have been studied with the EFT approach, and it has been indicating the possibility of NP explanations. 
 Based on the form factor in Ref.~\cite{Iguro:2020cpg}, we derive updated allowed regions for $C_{X}^{c\tau} (m_b)$ from the aforementioned latest experimental results,  assuming $C_{X}^{c\ell} (m_b)=0$.  
 The fit result for each NP scenario can be written as $C_{V_1}^{c\tau} (m_b) = +0.09(3)$, $C_{V_2}^{c\tau} (m_b) = \pm0.42(7) i$, and $\text{Re} C_{T}^{c\tau} (m_b) = +0.15(7)$ with $\text{Im} C_{T}^{c\tau} (m_b)$ to be fixed as $\pm 0.19$.
 We will also show allowed contour plots on the complex plane later (see Fig.~\ref{Fig:contour}) along with the LHC bound.
 Note that our present analysis excludes the scalar scenarios $S_{1,2}$. 
 In particular, the $S_{2}$ solution to the excesses is not consistent with the condition $\mathcal B(B_c \to \tau\nu) \lesssim 30\%$, extrapolated from the $B_c$ lifetime~\cite{Alonso:2016oyd}.\footnote{
 As known well, the $S_{1}$ scenario has no solution to the excesses, which results in more than $99.8\%$ CL exclusion.} \\[-0.7em]
 \item 
 $b \to u \tau\nu$: 
 there is the NP study for $\bar B \to \pi \tau\bar\nu$ and $\bar B \to \tau\bar\nu$ in Ref.~\cite{Tanaka:2016ijq}.
 We update the bounds as   
 $C_{V_1}^{u\tau} (m_b) = +0.03(15)$, $C_{V_2}^{u\tau} (m_b) = +0.02(15)$, $C_{S_{1/2}}^{u\tau} (m_b) = 0.00(4)$, $\mp0.53(4)$, and $C_{T}^{u\tau} (m_b) = +0.17(25)$, $-0.94(29)$, 
 which are zero-consistent within $1\sigma$, although the latter two have degenerated results.
\end{itemize} 
In addition to them, we also evaluate NP constraints from $B_c \to \ell\nu$ and $B \to \ell\nu$. 
The former fills missing pieces of the $C_{S_{1,2}}^{c\ell}$ constraints for $b \to c\ell\nu$, while the latter gives the flavor bound for $b \to u\ell\nu$, which is not shown above. 
Note that these two body decays are not affected by the $T$ operator because of the Lorenz structure. 
\begin{itemize} 
 \item
 The way to derive $\mathcal B(B_c \to \tau\nu) \lesssim 30\%$ from the $B_c$ lifetime is indeed independent of the lepton flavor.
 Therefore, the same condition can be employed to constrain the NP effect in $b \to c\ell\nu$.
Given the condition, the $S_{1,2}$ contributions are constrained as $| C_{S_{1,2}}^{ce, c\mu} (m_b) | < 0.8$ in the electron and muon modes.
 For both lepton modes, the $V_{1,2}$ bounds are very loose.\footnote{
 It is obtained as $| C_{V_{1,2}}^{c\ell} (m_b) | < \mathcal O(100)$. 
 We treat such a case as ``unbound'' in this paper.
 }  \\[-0.7em]
 \item
 The branching ratio of $B \to \mu\nu$ has been measured for the first time, 
 and the result is given as $\mathcal B (B \to \mu\nu) = (6.46 \pm 2.22 \pm 1.60) \times 10^{-7}$ ~\cite{Sibidanov:2017vph}. 
 Then, we can obtain $C_{V_{1/2}}^{u\mu} = \pm 0.2(4)$ and $C_{S_{1/2}}^{u\mu} = \pm 3(6) \times 10^{-3}$,  $\mp 35(6) \times 10^{-3}$. 
 On the other hand, the upper limit of the branching ratio for $B \to e\nu$ is so far obtained as $< 9.8 \times 10^{-7}$ (90\% CL), which can be compared with the SM value, $(9.8 \pm 1.4) \times 10^{-12}$. 
 Even in this case, the $S_{1,2}$ scenarios are restricted as $| C_{S_{1,2}}^{ue} | < 0.02$. 
\end{itemize} 
With the help of these evaluations, we have the exhaustive list of the flavor bounds as in Table~\ref{Tab:flavorbound}. 
These existing and newly evaluated flavor bounds will be compared with the LHC bounds obtained in this work. 
%
\begin{table}[t]
\renewcommand{\arraystretch}{1.3}
  \begin{center}
  \scalebox{0.8}{
  \begin{tabular}{lccccc}
  \hline\hline
  				& $V_1$ 		& $V_2$ 					& $S_1$ 		& $S_2$ 		& $T$ \\
  \hline
  $C_X^{c e}$ 		& -- 	& $0.02(1)$ 				& $|<0.8|$ 	& $|<0.8|$ 	& $0.02(1)$  \\
  \hline
   $C_X^{c \mu}$ 	& -- 	& $0.02(1)$			 	& $|<0.8|$ 	& $|<0.8|$ 	& $0.02(1)$   \\
  \hline
  $C_X^{c \tau}$ 	& $0.09(3)$ & $\pm0.42(7)i$ 		& excluded 	& excluded 	& $0.15(7)^{**}$  \\
  \hline
  $C_X^{u e}$ 		& -- 	& --		 				& $|<0.02|$ 	& $|<0.02|$ 	& -- \\
  \hline
   $C_X^{u \mu}$ 	& $0.2(4)$ 	&  $-0.2(4)$ 		&  $3(6) \!\times\! 10^{-3} \,^*$ 	&  $-3(6) \!\times\! 10^{-3} \,^*$ & -- \\
  \hline
  $C_X^{u \tau}$ 	& $0.03(15)$ 	&  $0.02(15)$		& $0.00(4)^*$ 	&  $0.00(4)^*$  & $0.17(25)^*$ \\
  \hline\hline
  \end{tabular} 
 }
  \caption{   
 Status of applicable flavor bounds. 
 ``$|< n|$'' means $|C_X| < n$ where those for $C_{S_{1,2}}^{ce,c\mu}$ are the bounds from the theoretical estimates whereas that for $C_{S_{1,2}}^{ue}$ is obtained from the $90\%$ CL upper limit.
 The results with $\!\,^*$ indicate that there exist another best fit point outside of $C_X \sim 0$. 
 The $C_T^{c \tau}$ result $(^{**})$ is given real part by fixing the imaginary part to be $\text{Im} C_T^{c \tau} = \pm 0.19$. 
 See the main text for these details.  
  }
  \label{Tab:flavorbound}
  \end{center}
\end{table}

\section{\boldmath EFT realizations: leptoquark models }
\label{sec:LQ}

At collider physics, on the other hand, the EFT description is not always applicable. 
It rather depends on details of the NP model and of the analysis to be used. 
In this work, we will test the LQ models to see at which scale the EFT limit of Eq.~\eqref{Eq:effH} becomes a good approximation for the present and future LHC searches.

The LQ interactions are classified in Ref.~\cite{Buchmuller:1986zs} with the general $SU(3)_c \times SU(2)_L \times U(1)_Y$ invariant form. 
In this work, we leave details of the model constructions, and then just consider the explicit interactions only relevant for the present study. 
In the following subsections, we introduce LQ interactions that generate each operator in Eq. (\ref{Eq:effH}). 
The SM gauge representations for the LQ fields are summarized in Appendix~\ref{sec:LQrepresentation}.

\subsection{\boldmath ${V_1}$ operator}
\label{sec:CV1}
The ${V_1}$ operator is constructed by the vector leptoquark $\text{U}_1^\mu$.
The interaction term of interest is written as  
\begin{align}
 \label{eq:V1op}
 \mathcal L_{V_1} = h_\text{LQ}^{ij} \Big(\bar u_i \gamma_\mu P_L \nu_j + \bar d_i \gamma_\mu P_L \ell_j \Big) \text{U}_{1}^\mu + \text{h.c.},
\end{align}
and then the corresponding WC for $b \to q_m \ell_n\bar\nu_n$ is obtained as  
\begin{align}
 \label{eq:LQV1}
 &2\sqrt{2}G_FV_{qb} C_{V_1}^{q_m\ell_n} = + { h_{\text{LQ}}^{mn} h_{\text{LQ}}^{*3n} \over M_{\text{LQ}}^2 } \,,& 
\end{align}
for $q_{1,2} =(u, c)$ and $\ell_{1,2,3}=(e,\mu,\tau)$ by performing the Fierz transformation. 
A similar realization for $C_{V_1}$ is possible by means of the other LQs of $\text{S}_1$, ${\mathbf S}_3$, and ${\mathbf U}_3^\mu$ as shown in Ref.~\cite{Sakaki:2013bfa}. 
See also Appendix~\ref{sec:LQrepresentation}.

\subsection{\boldmath ${V_2}$ operator}
\label{sec:CV2}
The ${V_2}$ operator is constructed by the scalar leptoquark $\text{R}_2^{2/3}$, where $2/3$ denotes the electromagnetic charge. 
The interaction term 
\begin{align}
 \label{eq:V2op}
 \mathcal L_{V_2} = \left( h_{\text{LQ}_1}^{ij} \bar u_{i} P_L \nu_{j} +  h_{\text{LQ}_2}^{ij} \bar d_{i} P_L \ell_{j} \right) \text{R}_2^{2/3} + \text{h.c.} \,, 
\end{align} 
leads to 
\begin{align}
 \label{eq:LQV2}
 &2\sqrt{2}G_FV_{qb} C_{V_2}^{q_m\ell_n} = + { h_{\text{LQ}_1}^{mn} h_{\text{LQ}_2}^{*3n} \over 2M_\text{LQ}^2 } \,.& 
\end{align} 
Concerning a practical model, we need two $SU(2)$ doublet LQ fields $\text{R}_2 = (\text{R}_2^{5/3}~\text{R}_2^{2/3})$ and $\text{R}'_2 = (\text{R}_2^{\prime 2/3}~\text{R}_2^{\prime -1/3})$ to construct the SM gauge invariant form, 
as written in Appendix~\ref{sec:LQrepresentation}. 
Since the LHC signature for the current process is unchanged, we do not consider such a case.

\subsection{\boldmath ${S_1}$ operator}
\label{sec:CS1}
The ${S_1}$ operator is constructed by the vector leptoquark $\text{U}_1^\mu$ such that 
\begin{align}
 \label{eq:S1op}
 \mathcal L_{S_1} = \left( h_{\text{LQ}_1}^{ij} \bar u_{i} \gamma_\mu P_L \nu_{j} + h_{\text{LQ}_2}^{ij} \bar d_{i} \gamma_\mu P_R \ell_{j} \right) \text{U}_{1}^\mu + \text{h.c.} \,, 
\end{align} 
and we have 
\begin{align}
 \label{eq:LQS1}
 &2\sqrt{2}G_FV_{qb} C_{S_1}^{q_m\ell_n} = -{2 h_{\text{LQ}_1}^{mn} h_{\text{LQ}_2}^{*3n} \over M_\text{LQ}^2 }\,.& 
\end{align}
Another realization is given by the $\text{V}_2^\mu$ LQ~\cite{Sakaki:2013bfa}.

\subsection{\boldmath ${S_2}$ and $T$ operators}
\label{sec:CS2}
The ${S_2}$ and $T$ operators are always connected due to property of the Fierz transformation. 
They are constructed by the two scalar leptoquarks $\widetilde{\text{R}}_2^{2/3}$ and $\widetilde{\text{S}}_1$. 
Note that for $\widetilde{\text{LQ}}$ index of the (quark, lepton) pair is flipped as (lepton, quark) in our notation just for our calculating convenience. 
The single ${S_2}$ and $T$ terms are respectively realized from 
\begin{align}
 \label{eq:S2op}
  \mathcal L_{S_2} = 
 & \left(-\tilde h_{\text{LQ}_1}^{ji} \bar \nu_{j}  P_R d_{i}^c + \tilde h_{\text{LQ}_2}^{ji} \bar \ell_{j}  P_L u_{i}^c \right) \widetilde{\text{S}}_1 \notag \\
 & +\left( \tilde h_{\text{LQ}_2}^{ji} \bar \nu_{j} P_R u_{i} - \tilde h_{\text{LQ}_1}^{ji} \bar \ell_{j} P_L d_{i} \right) \widetilde{\text{R}}_2^{2/3} + \text{h.c.} \,, \\
 \label{eq:LQS2}
 & \text{with}~~ 2\sqrt{2}G_FV_{qb} C_{S_2}^{q_m\ell_n} = - { \tilde h_{\text{LQ}_1}^{n3*} \tilde h_{\text{LQ}_2}^{nm} \over M_\text{LQ}^2 } \,,
\end{align} 
and 
\begin{align}
 \label{eq:Top}
  \mathcal L_{T} = 
 & \left(-\tilde h_{\text{LQ}_1}^{ji} \bar \nu_{j}  P_R d_{i}^c + \tilde h_{\text{LQ}_2}^{ji} \bar \ell_{j}  P_L u_{i}^c \right) \widetilde{\text{S}}_1 \notag \\
 & -\left( \tilde h_{\text{LQ}_2}^{ji} \bar \nu_{j} P_R u_{i} + \tilde h_{\text{LQ}_1}^{ji} \bar \ell_{j} P_L d_{i} \right) \widetilde{\text{R}}_2^{2/3} + \text{h.c.} \,, \\
 \label{eq:LQT}
 & \text{with}~~ 2\sqrt{2}G_FV_{qb} C_{T}^{q_m\ell_n} = + { \tilde h_{\text{LQ}_1}^{n3*} \tilde h_{\text{LQ}_2}^{nm} \over 4M_\text{LQ}^2 } \,,
\end{align}
where $u^c$ and $d^c$ denote charge conjugated quarks.
Again, a practical model requires the $SU(2)$ doublet LQ field $\text{R}_2 = (\text{R}_2^{5/3}~\text{R}_2^{2/3})$.

\subsection{\boldmath Mass scale restriction}
\label{sec:MassScale}

The LQ searches have been given by the QCD pair production and the single production channels with subsequent decays into a pair of quark and lepton. 
Then the recent studies~\cite{Sirunyan:2018jdk,Aaboud:2019bye,Aad:2020iuy} obtain the lower limit on the LQ mass as $\sim 1.5~\text{TeV}$ by assuming 100\% branching ratio for the subsequent decay.

On the other hand, the present high $p_T$ searches with the $l + \text{missing}$ events can access a larger LQ mass region since LQs produce non-resonant signals in this case.

\begin{figure}[t!]
\begin{center}
\includegraphics[viewport=0 0 620 311, width=26em]{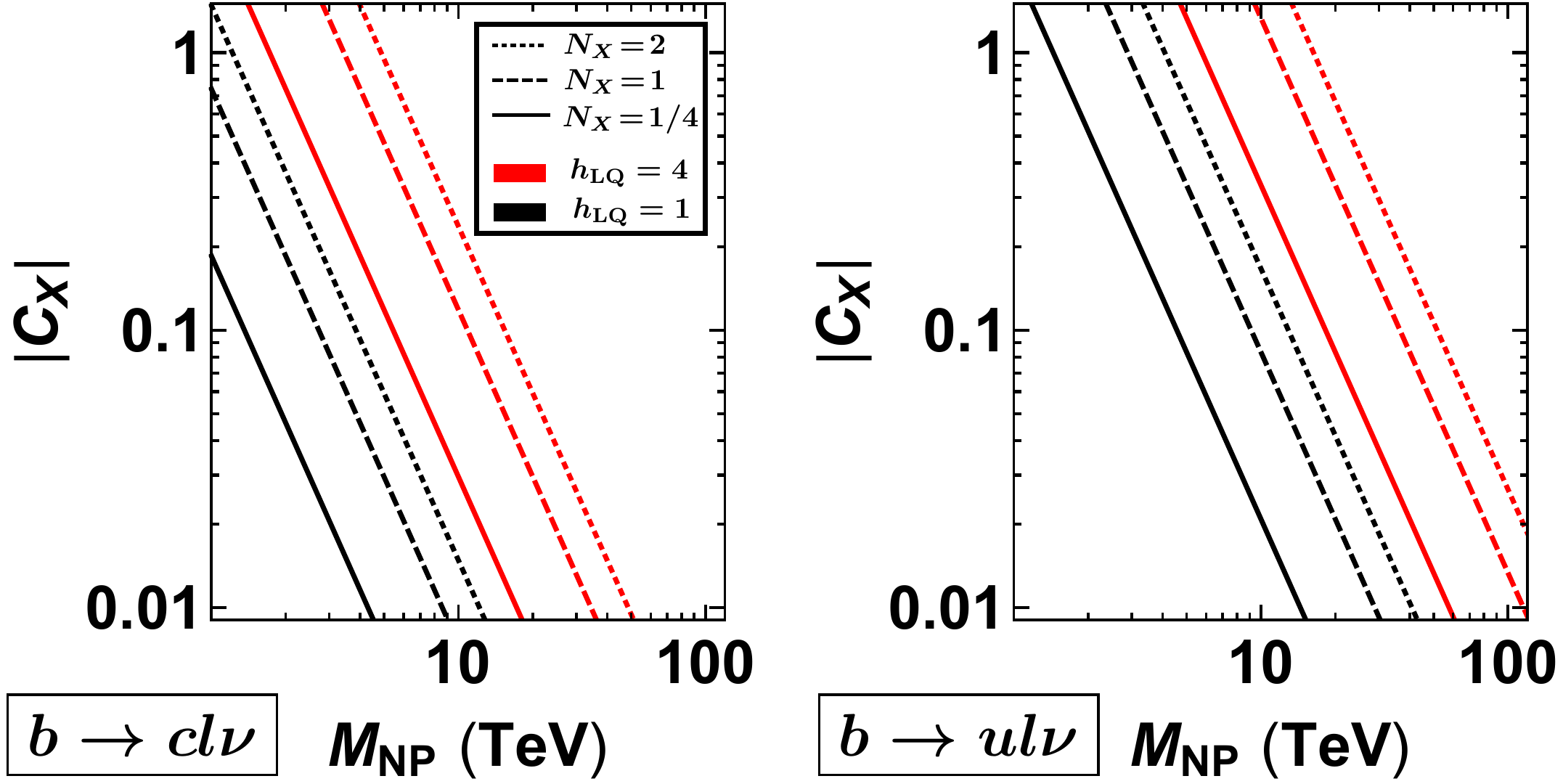}
\caption{
\label{Fig:MassScaleCheck}
Relations between the LQ mass and the WC at the NP scale by fixing the LQ coupling. 
} 
\end{center}
\end{figure}

Once $C_X$, at any scale, is given from flavor observables and/or LHC studies, the corresponding LQ mass is restricted as long as the LQ coupling is set not too large for perturbation theory \cite{DiLuzio:2017chi}. 
In Fig.~\ref{Fig:MassScaleCheck}, we plot the relation between $C_X (\Lambda)$ and $M_\text{NP}$ of Eq.~\eqref{eq:WCEFT}. 
The numerical factor $N_X$ is fixed as explained in the legend. 
The Lorenz structures correspond to $N_{S_1}=2$; $N_{V_1,S_2} = 1$; $N_{V_2}=1/2$; and $N_T=1/4$ for the present LQ models. 
Then, one can check the accessible LQ mass from the plot.  
For instance, $|C_X^{cl} (\Lambda)| \sim \mathcal O(0.01)$ is produced with $M_\text{LQ} \lesssim 10\,\text{TeV}$ for $h_\text{LQ} \lesssim 1$ in $ b \to c l \nu$. 
The mass could be $M_\text{LQ} \sim 50\,\text{TeV}$ at maximum if we allow the coupling as large as $h_\text{LQ} =4$.

\section{\boldmath Collider study}
\label{sec:collider}

At the LHC, the EFT operators of Eq.~\eqref{Eq:effH} contribute to $pp \to l^\pm + \text{missing}$ from $b \bar q \to l^-\bar\nu$ and $\bar b q \to l^+\nu$ for $q = u, c$. 
The SM contribution is dominantly given by the $W^\pm$ exchange and generates a resonant structure at $M_{W^{\pm}}$ in the distribution for the transverse mass 
\begin{align}
 m_T = \sqrt{2p_\text{T}^l E_\text{T}^\text{miss} (1-\cos \phi_{l\nu})}  \,,
\end{align} 
where $\phi_{l\nu}$ is the azimuthal angle between $l$ and the missing momentum.
On the other hand, the present NP effects are off resonant and widely spread in the $m_T$ distribution. 
Thus one can expect that a tail of the resonance, namely a larger $m_T$ range, is sensitive to the NP effect.

\subsection{\boldmath EFT limit}
\label{sec:Xsl}

To see NP effects on signal event distributions, here we show analytic forms of the parton-level cross sections for $b \bar c \to e^- \bar\nu$ in the LQ models, 
and then see the EFT limit of the models. 
Defining the four momenta as 
\begin{align}
 p_c^\mu & = E(1, 0, 0, 1) \,, \label{Eq:pb} \\ 
 p_b^\mu & = E(1, 0, 0, -1) \,, \label{Eq:pc} \\
 p_e^\mu & = E(1, \sin\theta, 0, \cos\theta) \,, \\
 p_\nu^\mu & = E(1, -\sin\theta, 0, -\cos\theta) \,, 
\end{align}
we obtain 
\begin{align}
 { d\sigma_X \over d\cos\theta } = {1 \over 3} {|\mathcal M_X|^2 \over 128\pi E^2} \,, 
\end{align}
with the spin averaged sum of the squared matrix element, namely $|\mathcal M_X|^2$ (${1 \over 4}\sum_\text{spin}$ is abbreviated), written as 
\begin{align}
  | \mathcal M_{V_1}^\text{LQ} |^2 = 
  & \,4\, (h_\text{LQ}^{21} h_\text{LQ}^{31*})^2 E^4 \hat C_{t}^2 (1 - \cos\theta)^2 \,, \label{Eq:XsV1} \\[0.5em]
 | \mathcal M_{V_2}^\text{LQ} |^2 = 
 & \,(h_{\text{LQ}_1}^{21} h_{\text{LQ}_2}^{31*})^2 E^4 \hat C_{t}^2 (1 + \cos\theta)^2 \,, \label{Eq:XsV2} \\[0.5em]
 | \mathcal M_{S_1}^\text{LQ} |^2 = 
 & \,16\, ( h_{\text{LQ}_1}^{21} h_{\text{LQ}_2}^{31*})^2 E^4 \hat C_{t}^2 \,, \\[0.5em]
 | \mathcal M_{S_2/T}^\text{LQ} |^2 = 
 & \,( \tilde h_{\text{LQ}_2}^{12*} \tilde h_{\text{LQ}_1}^{13} )^2 E^4 \left[ \hat C_{t}^2 (1 + \cos\theta)^2 \right. \label{Eq:XsS2T} \\
 &\left.+ \hat C_{u}^2 (1 - \cos\theta)^2 \pm 2 \hat C_{t} \hat C_{u} (1-\cos^2\theta) \right] \,, \notag
\end{align}
where $\hat C_{t}$ and $\hat C_{u}$ involve the LQ propagator written as 
\begin{align}
 \label{eq:Prop_t}
 \hat C_{t} & = \left[ 2 E^2 (1 + \cos\theta) + M_\text{NP}^2 \right]^{-1} \,, \\
 \label{eq:Prop_u}
 \hat C_{u} & = \left[ 2 E^2 (1 - \cos\theta) + M_\text{NP}^2 \right]^{-1} \,. 
\end{align}
At the EFT limit, the angular and energy dependence of the propagator is suppressed such that $\hat C_{t} \simeq \hat C_{u} \simeq 1/M_\text{NP}^2$, and thus we have 
\begin{align}
  | \mathcal M_{V_1}^\text{LQ} |^2 \simeq 
  & \,4\, { (h_\text{LQ}^{21} h_\text{LQ}^{31*})^2 \over M_\text{LQ}^4} E^4 (1 - \cos\theta)^2 \,, \\
 | \mathcal M_{V_2}^\text{LQ} |^2 \simeq
 & \, { (h_{\text{LQ}_1}^{21} h_{\text{LQ}_2}^{31*})^2 \over M_\text{LQ}^4} E^4 (1 + \cos\theta)^2 \,, \\
 | \mathcal M_{S_1}^\text{LQ} |^2 \simeq 
 & \,16\, { (h_\text{LQ}^{21} h_\text{LQ}^{31*})^2 \over M_\text{LQ}^4} E^4 \,, \\
 | \mathcal M_{S_2}^\text{LQ} |^2 \simeq 
 & \,4\, {( \tilde h_{\text{LQ}_2}^{12*} \tilde h_{\text{LQ}_1}^{13} )^2 \over M_\text{LQ}^4} E^4 \,, \\
 | \mathcal M_{T}^\text{LQ} |^2 \simeq
 & \,4\, {( \tilde h_{\text{LQ}_2}^{12*} \tilde h_{\text{LQ}_1}^{13} )^2 \over M_\text{LQ}^4} E^4 \cos^2\theta \,. 
\end{align}
We can see that the relations of Eqs.~\eqref{eq:LQV1}, \eqref{eq:LQV2}, \eqref{eq:LQS1}, \eqref{eq:LQS2}, \eqref{eq:LQT} are achieved in this limit.

The parton energy $E$ fluctuates in the proton, and thus it is distributed as $0 < E < \sqrt{s}/2$. 
The energy distribution is weighted with Parton Distribution Function (PDF) according to which $b$ and $\bar c$ quarks tend to have low energy. 
Namely, the distribution in a high energy range is suppressed, and hence the EFT limit is a good approximation even for $M_\text{NP} \lesssim \mathcal O(1) \,\text{TeV}$ as long as the total cross section is concerned.

However, this is not the case for our analysis. 
For the present process, the NP sensitivity is gained at large $p_T$ due to the large SM background at lower $p_T$. 
In other words, the LHC bound for NP is provided from the signal event distribution with high $p_T$, which arises from the high energy parton at the cost of the PDF suppression. 
To be precise, $m_T \sim 2\,\text{--}\,3\,\text{TeV} (\sim\,E)$ is the most sensitive for the present case. 
Therefore, the EFT limit $E^2 \ll M_\text{NP}^2$ should be valid only for $M_\text{NP} > \mathcal O(10) \,\text{TeV}$.

If $E^2 \lesssim M_\text{NP}^2$ is in the case, the angular and energy dependence in the propagator $\hat C_{t}$ and $\hat C_{u}$ is of critical importance since it affects the $m_T$ distribution,
as will be seen below.

\subsection{\boldmath Numerical analysis}
\label{sec:analysis}

Here we perform numerical analyses to obtain constraints on the WCs from the $pp \to l^\pm + \text{missing}$ searches at the 13~TeV LHC by means of the aforementioned LQ models described in Sec.~\ref{sec:LQ}. 
In this work, we apply numerical setup of the $\ell\nu$ and $\tau\nu$ searches by ATLAS~\cite{Aad:2019wvl} and CMS~\cite{Sirunyan:2018lbg}, respectively.

\subsubsection{simulation setup}
\label{sec:setup}

We generate 100K signal events for every LQ mass $M_\text{LQ} = 2$, $3$, $5$, $7.5$, $10$, $20$, and $100\,\text{TeV}$ in each model, 
by using {\tt Madgraph5}~\cite{Alwall:2014hca} with {\tt NNPDF2.3}~\cite{Ball:2012cx} through {\tt PYTHIA8}~\cite{Sjostrand:2014zea} within the five flavor scheme. 
These generated events are interfaced to {\tt DELPHES3}~\cite{deFavereau:2013fsa} for the fast detector simulation. 
Then, we apply the following sets of the selection cuts. 
For the $\ell=e,\mu$ modes, we require
$n_{\ell} = 1$, $p_{T,\ell}> 65~\GeV$, $|\eta_{\ell}|> 2.5$, 
and $E\!\!\!/_T > 55~\GeV$ ($\mu$ mode), $65~\GeV$ ($e$ mode)
following the ATLAS search with 139~fb$^{-1}$ at 13~TeV as in Ref.~\cite{Aad:2019wvl}.
Regarding the $\tau$ mode, we require exactly one hadronically decaying tau, 
$n_{\tau_h} = 1$, with $p_{T,\tau} > 80~\GeV$ and $|\eta_\tau| < 2.1$, 
no isolated $e$ or $\mu$, 
$E\!\!\!/_T > 200~\GeV$, $0.7< p_{T,\tau}/E\!\!\!/_T < 1.3$, 
and $|\Delta \phi(p_{\tau},E\!\!\!/_T)|>2.4$, 
following the CMS search with 35.9~fb$^{-1}$ at 13~TeV as in Ref.~\cite{Sirunyan:2018lbg}.

\begin{figure}[t!]
\begin{center}
\includegraphics[viewport=0 0 567 568, width=20em]{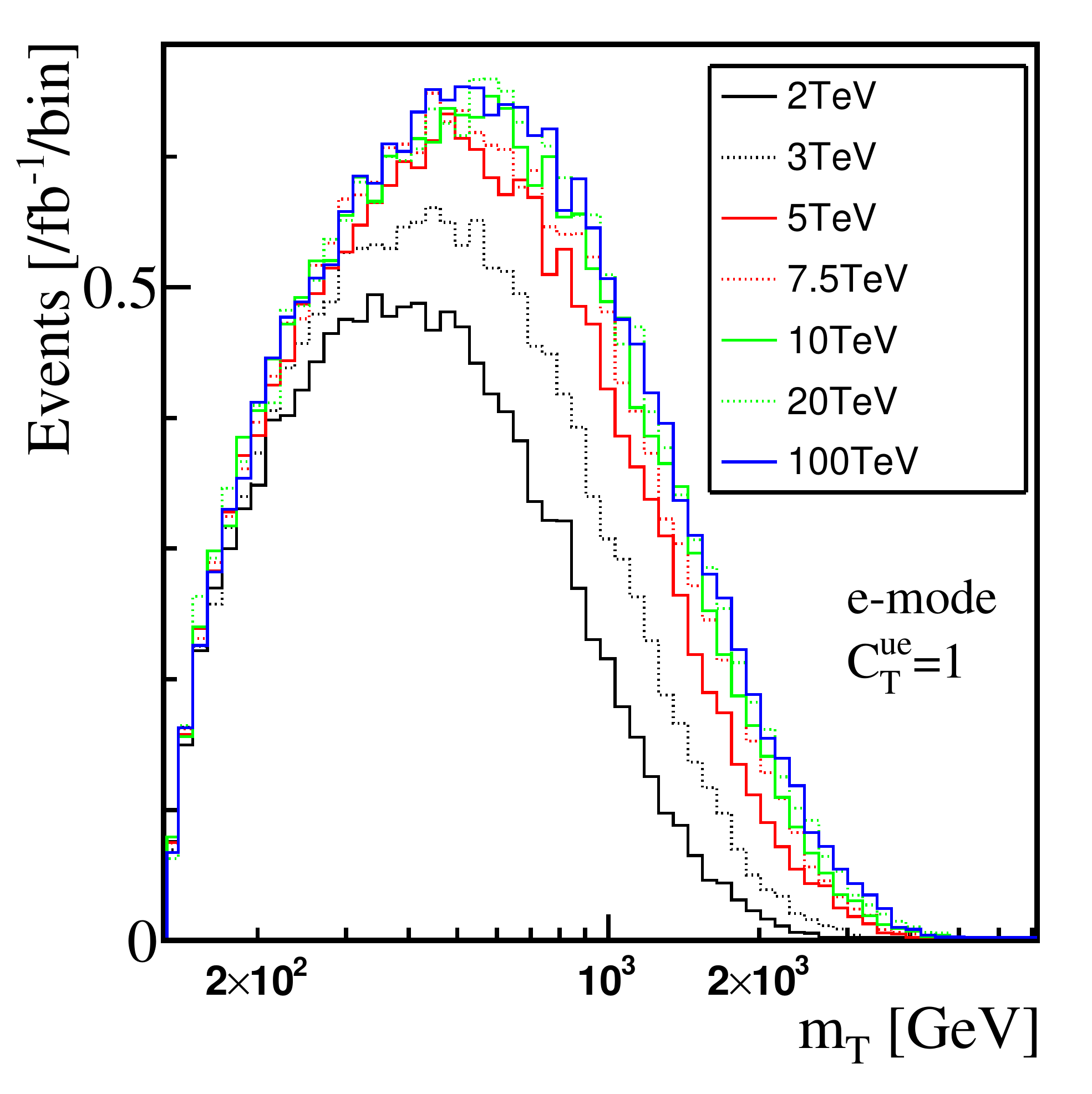}
\caption{
\label{Fig:Distributions}
The simulated $m_T$ distributions of $pp \to e^\pm + \text{missing}$ for the tensor type NP in the $bu\to e\nu$ mode with $M_\text{LQ}=2, 3, 5, 7.5, 10, 20$ and $100\,\text{TeV}$ by fixing $C_T^{ue} (\Lambda_\text{LHC})=1$.
} 
\end{center}
\end{figure}

Figure~\ref{Fig:Distributions} shows the $m_T$ distributions of $pp \to e^\pm + \text{missing}$ for the tensor type NP in the $bu\to e \nu$ mode 
by fixing $C_T^{ue} (\Lambda_\text{LHC})$ $=1$, but with different values of $M_\text{LQ}$ as an illustration. 
One can see that the distributions converge at the low $m_T$ region, which implies that the EFT limit is valid as it is for the flavor phenomenology. 
On the other hand, the discrepancy among them at the high $m_T$ region is significant.
When the WC is fixed, the more events are expected as $M_\text{LQ}$ increases, 
which is because the relative importance of the energy dependent terms in the $t/u$-channel propagators in Eqs. (\ref{eq:Prop_t}) and (\ref{eq:Prop_u}) is larger for the same $m_T$ value.
Similar tendencies are observed for all the type of operators, and also for the $bc\to l\nu$ cases.
Regarding the $bc\to l\nu$ cases, however, the size of the discrepancy among the masses are relatively small, 
since the initial parton contributing to those processes is less energetic $c$-parton than $u$-parton.
Nevertheless, it generates a significant difference in the LHC bound.

\subsubsection{LHC bound on $C_{X} (\Lambda_\text{LHC})$}
\label{sec:bound}

After the event selection cuts, we obtain the $m_T$ distributions and extract the constraints on the WCs based on them.
For the $e, \mu$ modes, we use the $95\%$ confidence level (CL) upper bounds on the model independent NP contributions of $m_T > m_{T,\min}$, 
provided in Table 4 and Table 5 of Ref.~\cite{Aad:2019wvl}.
We take the $m_{T,\min}$ threshold value providing the strongest constraint for each model. Note that for the electron mode, 
a deficit of the events in the tale region is observed, which results in stronger constraints than expected.
For the $\tau$ mode, we perform the same analysis based on the background $m_T$ distribution in Ref.~\cite{Sirunyan:2018lbg}.
To be conservative, we assigned a $30$\,--\,$50\%$ systematic uncertainty for $m_{T,\min}=$\,1.5\,--\,3~TeV.

Then, an excluded upper limit on the LQ coupling(s) $h_{\text{LQ}_{(i)}}$ is derived for every $M_\text{LQ}$. 
Finally, we translate it into $C_X (\Lambda_\text{LHC})$. 
In this work, we fix the LHC scale to be $\Lambda_\text{LHC} = 1\,\text{TeV}$ for simplicity. 
Note that we found that the present data is sensitive to the considering NP signal events in the region of $m_T \sim 2\,\text{--}\,3\,\text{TeV}$.

\begin{figure*}[t!]
\begin{center}
\includegraphics[viewport=0 0 972 432, width=46em]{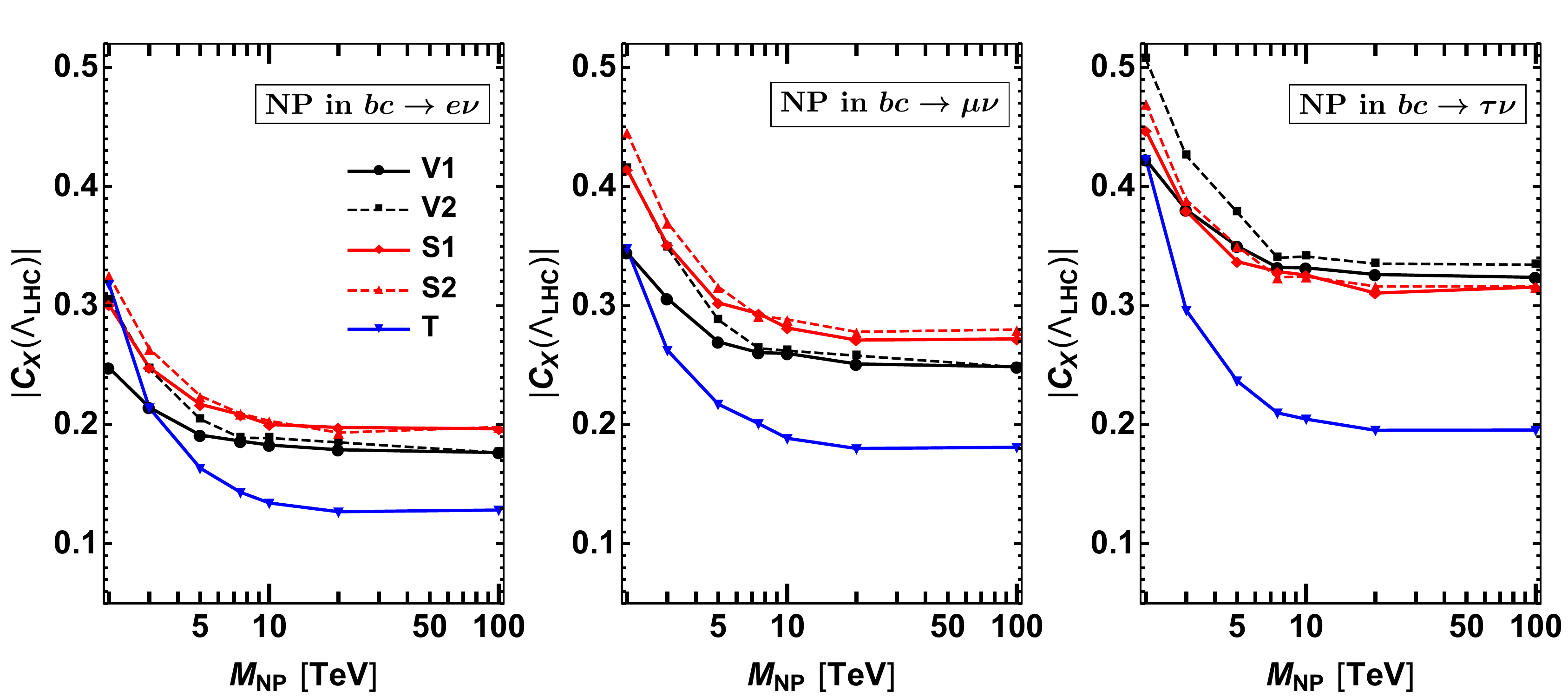} \\
\includegraphics[viewport=0 0 972 432, width=46em]{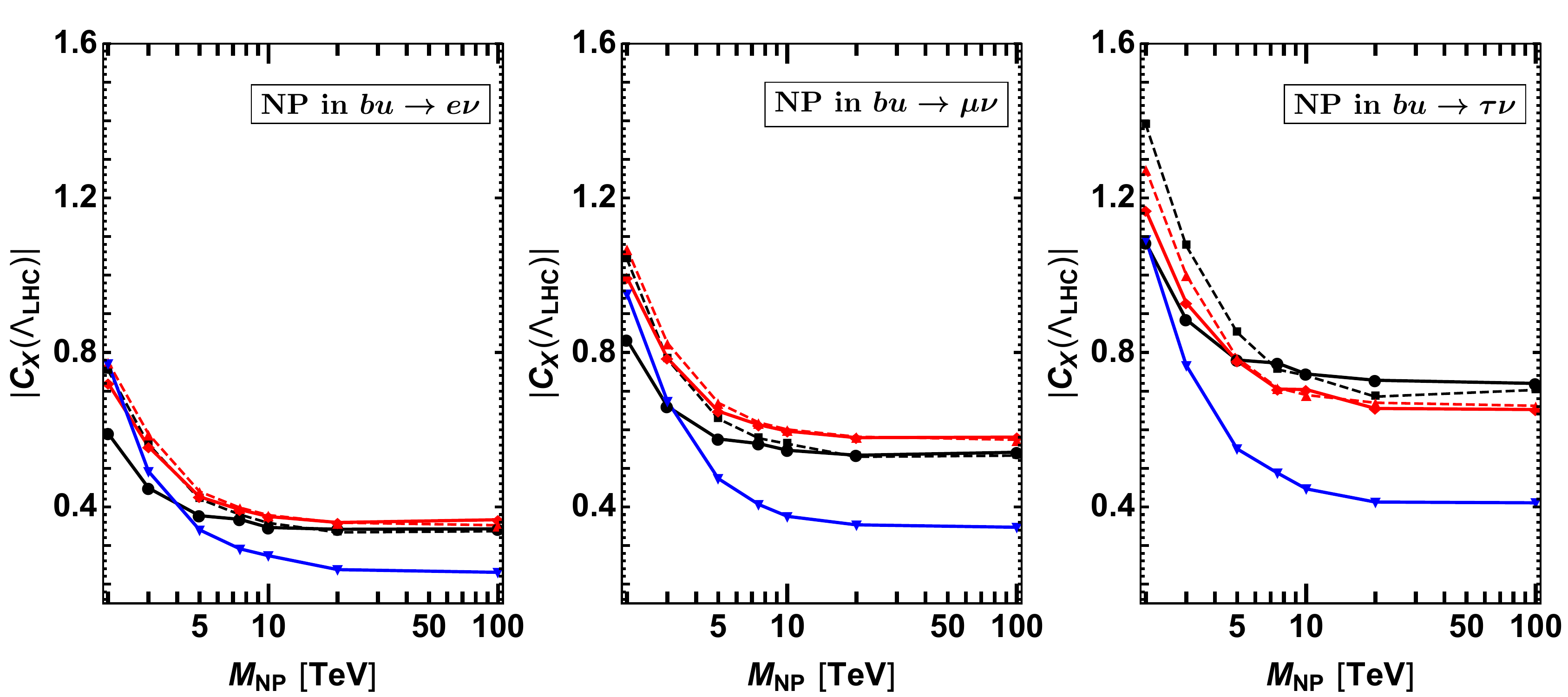} 
\caption{
\label{Fig:LHCbound}
The 95\% CL upper bounds on $C_X^{ql} (\Lambda_\text{LHC})$ obtained from the $\ell\nu$ and $\tau\nu$ search data by ATLAS~\cite{Aad:2019wvl} and CMS~\cite{Sirunyan:2018lbg}, respectively, 
where we fix $\Lambda_\text{LHC} = 1\,\text{TeV}$.
} 
\end{center}
\end{figure*}

In Fig.~\ref{Fig:LHCbound}, we show the upper bounds on $C_X^{ql} (\Lambda_\text{LHC})$ at 95\% CL with respect to the fixed $M_\text{NP}$ for all the combinations of $X = (V_1$, $V_2$, $S_1$, $S_2$, $T)$, $q=(c,u)$, and $l = (e, \mu, \tau)$. 
One can clearly see that our results of the LHC bounds depend on the LQ mass for the region of $M_\text{LQ} < 10\,\text{TeV}$, while not for the larger LQ mass, as expected. 
In particular, the lower LQ mass results in the milder LHC bound on the WCs,\footnote{
For the NP models with a $s$-channel mediator, {\it e.g.} a charged scalar or vector boson, the EFT description usually gives weaker bounds at the LHC, which is opposite to the present case. 
}
which is straightforwardly inferred from  Fig.~\ref{Fig:Distributions}. 
We also found that the mass dependence for the $T$ type NP is more significant than one for the other types of NP. 
This is a non-trivial feature from the angular dependence as in Eq.~\eqref{Eq:XsS2T}.

One also finds that the EFT results ($M_\text{LQ} > 10\,\text{TeV}$) are independent of the chirality of the fermions, and only sensitive to the Lorentz structure. 
However, this does not hold when the EFT limit is not valid.  
This is because that the angular and energy dependence in the propagator $\hat C_{t,u}$ of Eqs.~\eqref{Eq:XsV1} -- \eqref{Eq:XsS2T} affects the $m_T$ distribution.

One may be interested in the LQ scenario with respect to the single $\widetilde{\text{R}}_2$ ($\widetilde{\text{S}}_1$) LQ particle 
that generates the relation $C_{S_2}  = +4C_T$ ($C_{S_2} = -4C_T$) at the LQ scale in the EFT limit. 
Indeed, the differential cross section for the $\widetilde{\text{R}}_2$ ($\widetilde{\text{S}}_1$) scenario is easily derived from Eq.~\eqref{Eq:XsS2T} by replacing $\hat C_u (\hat C_t) \to 0$.  
Then, we can see that the expression coincides with Eq.~\eqref{Eq:XsV2} of the $V_2$ operator (Eq.~\eqref{Eq:XsV1} of $V_1$) by scaling the factor 1 (4) at the EFT limit. 
Thus, the LHC bounds of these two scenarios, for the $e$ and $\mu$ modes, can be read off from those of $C_{V_1}$ and $C_{V_2}$ in Fig.~\ref{Fig:LHCbound}. 
For the $\tau$ mode, on the other hand, the $\tau_L$/$\tau_R$ difference in the effective operators for $C_{V_{1,2}}$ and $C_{S_2}=\pm4C_T$ affects the analysis due to the $\tau$ decay property at the LHC~\cite{Papaefstathiou:2011kd}. 
We will see this point in Sec.~\ref{sec:discussion}.

\subsection{\boldmath Future prospects}
\label{sec:prospect}

In turn, we discuss future sensitivities for the NP searches from $pp \to l^\pm + \text{missing}$ at the high luminosity (HL)-LHC with 3 ab$^{-1}$ data. 
Re-scaling the BG information which is used to derive the current bounds including the BG uncertainties, 
we can obtain the prospect of the HL-LHC bounds on the WCs in Table~\ref{Tab:LHCsummary} denoted as ``(w/sys)''.  
In the table, the results for the $(2\,\text{TeV} \text{\,--\,} 100\,\text{TeV})$ mass of the LQ particle are summarized. 
Since the BG uncertainty in the tail region is significant, 
we also show the optimistic cases that only the statistic error is taken and the theory uncertainty is controlled (negligible) in future, which is given in ``(no~sys)'' rows.

Furthermore, since the SM background at the tail of the $m_T$ distribution is dominated by the $W^\pm$-contribution \cite{Sirunyan:2018lbg,Aad:2019wvl},
we test $S/B$ improvement by selecting $l^-$ events as explained below.
Since the luminosity functions for $u\bar{d}$ and $d\bar{u}$ are not identical but the ratio is $L(u\bar{d})/L(d\bar{u}) \gtrsim 4$ above $2\,\text{TeV}$, 
the $l^+$ events are more observed than $l^-$ in the SM. 
Thus, the ratio for the $l^+/l^-$ events would be expected as $N_{l^+}/N_{l^-}\gtrsim 4$ in the tail region for the BG contribution. 
On the other hand, no charge asymmetry is expected between the $c\bar{b}$ and $b\bar{c}$ cases, namely $L(c\bar{b})/L(b\bar{c}) \sim 1$. 
Therefore, selecting only the $l^-$ events would improve the significance for the $b \to c l\nu$ process.
The results by selecting $l^-$ events are given in the "(sys, $l^-$)" rows,
where we assume the BG contributions reduced to $1/4$ 
and the $S/B$ factor would be improved about twice.
It turned out that selecting only the $l^-$ events can potentially improve the sensitivity to $C_X^{cl}$ by $30\,\text{--}\,40\%$ as seen in the table.
We have also checked that the effect of the selecting $l^-$ events 
with the no BG systematic uncertainty. The results are slightly improved 
but almost the same numbers in the ``(no sys)" rows are obtained, thus, not shown in the table.
The reason is that in principle the selection cut does not improve $S/\sqrt{B}$ and neither does the resulting sensitivity, if already the BG uncertainty is well controlled.
In other words, the $S/B$ improvement by the selection cut 
minimizes the effect of the BG uncertainty.
For the case of the $b \to u l\nu$ process, the signal charge asymmetry turns out to be larger than that for the SM BG due to the large ratio of 
$L(u\bar{b})/L(\bar{u}b)$. 
Hence, selecting $l^+$ is efficient for this case, but the improvement would be limited.

In any case, we think that the charge asymmetry defined as $A_{l}= (N_{l^+}-N_{l^-})/(N_{l^+}+N_{l^-})$ would be 
useful for a more dedicated study in
$(m_T,A_{l})$ distribution.

If the systematic error is well controlled, 
the $m_T$ bin with a large number of events will determine the 
bounds, and thus 
the smaller $m_T$ bin will become more relevant.
On the other hand, when the systematic error is large,
the bins closer to the tail will be more 
effective to set the bounds, since the background number of 
events should be non-negative.
We found that 
the $m_{T,\min}$ value providing the strongest bound
lies in $2$\,--\,$3$~TeV even for the HL-LHC. 
Therefore, the mass dependence will remain important, as long as the systematic error is non-negligible.

The detailed statistical analysis procedures for the future prospects  are as follows.
For each threshold bin $i$, we compute the value of $S_{i}^{95}$ based on the Poisson 
distribution satisfying the following criteria.
\begin{align}
\int_0^{B^0_i} dB_i f(B_i) P(S_{i}^{95} + B_i, N_{i,{\rm BG}}) = 0.05,    
\end{align}
where $P(S, N)$ is the probability to observe $N$ or less based on Poisson distribution of $S$, 
and $f(B_i)$ is the probability distribution of $B_i$, the BG contribution for the 
threshold bin $i$, and taken as the Gaussian distribution with $B_i^0$ and $\Delta B_i$ 
restricted to be in the range of $0\leq B_i \leq B_i^0$, where the normalization is taken as $\int_0^{B^0_i} dB_i 
f(B_i)=1$. 
We take the observed number $N_{i, BG}$ in future 
as the most frequent value based on the BG distribution.
Based on the above procedure, the corresponding upper bound 
$C_X$ at 95\% CL for the each threshold bin $i$ is obtained. 
The minimum value among $i$ is taken as the 95\% CL upper bound of $C_X$.

Another possibility for future NP searches is to utilize the pseudo rapidity distribution. 
We discuss it at length in Appendix~\ref{sec:angulardistribution}.

\section{\boldmath Combined constraints on $C_{X} (m_b)$}
\label{sec:discussion}

Here we summarize all the constraints on the WCs at the low energy scale $\mu = m_b$ both from the LHC and flavor bounds evaluated in this work.

The RGE running effect of $C_X$ from $\Lambda_\text{LHC} = 1\,\text{TeV}$ to $m_b = 4.2\,\text{GeV}$ is numerically presented as
\begin{align}
\begin{pmatrix}
C_{V_1} (m_b) \\
C_{V_2} (m_b) \\
C_{S_1} (m_b) \\
C_{S_2}(m_b) \\
C_T (m_b)  
\end{pmatrix}
\simeq
\begin{pmatrix}
1 & 0 &0&0&0\\
0 & 1&0&0&0\\
0 & 0&1.71&0&0\\
0 & 0&0&1.71&-0.27\\
0 & 0&0&0&0.84
\end{pmatrix}
\begin{pmatrix}
C_{V_1} (\rm{\Lambda_{LHC}}) \\
C_{V_2} (\rm{\Lambda_{LHC}}) \\
C_{S_1} (\rm{\Lambda_{LHC}}) \\
C_{S_2} (\rm{\Lambda_{LHC}}) \\
C_T (\rm{\Lambda_{LHC}})  
\end{pmatrix} \,, 
\label{eq:RGE}
\end{align} 
independent of the $(ql)$ index by following the formula in Ref. \cite{Iguro:2018vqb}, (see, also Refs.  \cite{Jenkins:2013wua,Alonso:2013hga,Gonzalez-Alonso:2017iyc}). 
Hence, the LHC bounds of $C_X^{ql} (m_b)$ are easily obtained by rescaling our results in Fig.~\ref{Fig:LHCbound}. 
The $S_2$-$T$ mixing propagates $C_{S_2} (\rm{\Lambda_{LHC}})$ and $C_T (\rm{\Lambda_{LHC}})$ to $C_{S_2} (m_b)$.

Regarding the flavor bounds, we derived the updated values as written in Sec.~\ref{sec:EFT}
with the use of the recent experimental data and theoretical input~\cite{Amhis:2019ckw,Zyla:2020zbs,Aoki:2019cca}.

\begin{figure*}[t!]
\begin{center}
\includegraphics[viewport=0 0 1080 386, width=48em]{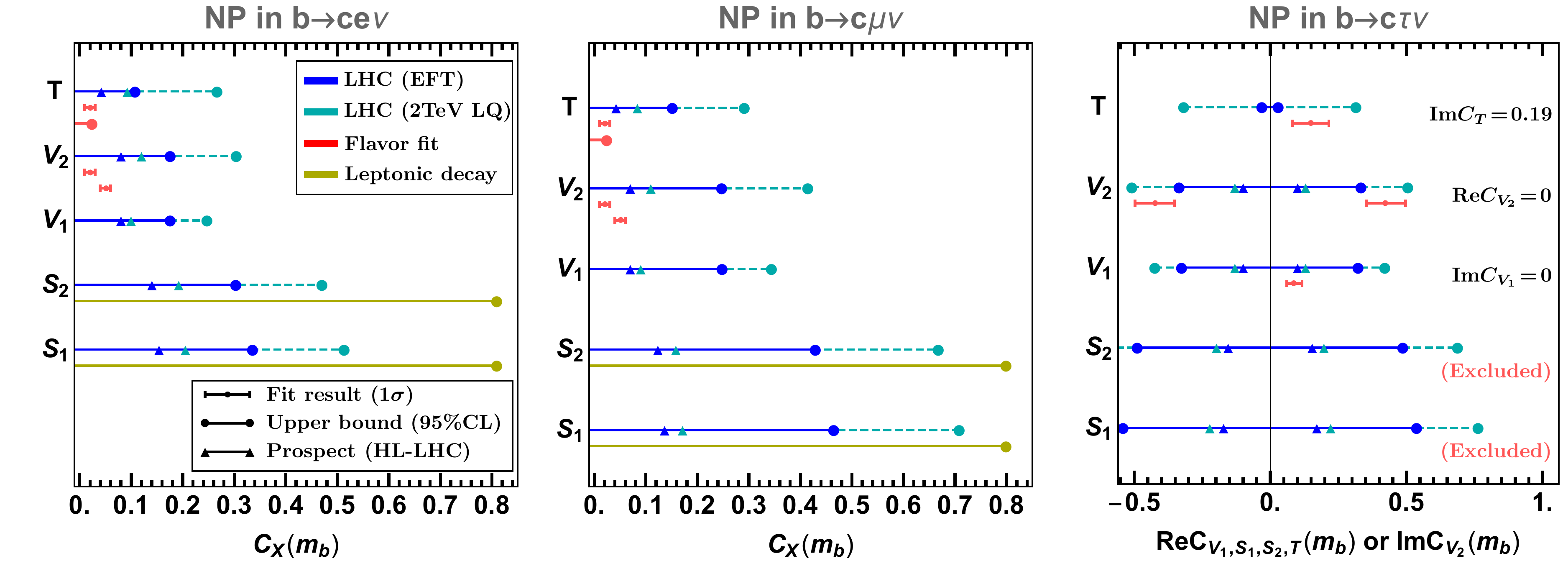} \\[0.5em]
\includegraphics[viewport=0 0 1080 386, width=48em]{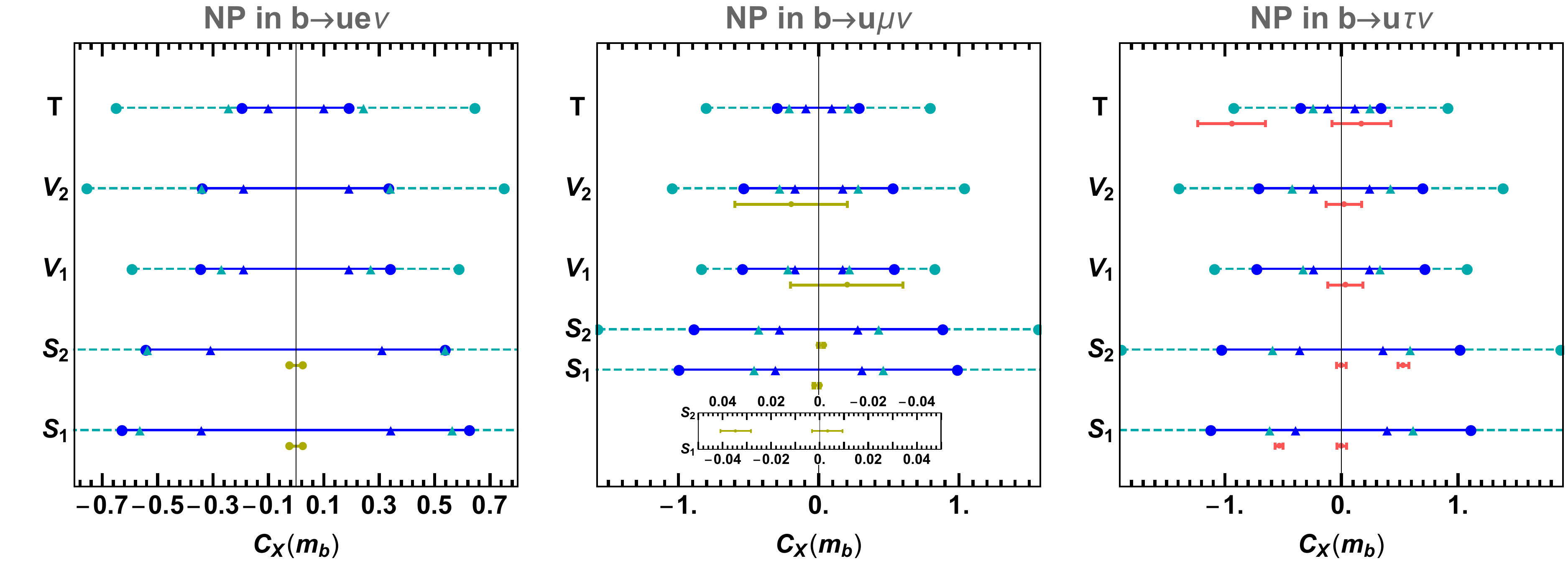} 
\caption{
\label{Fig:combined}
Summary of the flavor and LHC bounds of the WCs at $\mu = m_b$ along with the future prospects at the HL-LHC with $3\,\text{ab}^{-1}$ from Table~\ref{Tab:LHCsummary}. 
Unless the flavor bound indicates a favored direction on the complex plane, the WC is taken as real. 
Note that the prospect for $\text{Re}\, C_T^{c\tau}$ cannot be drawn since the assumption $\text{Im}\, C_T^{c\tau} =0.19$ already exceeds the result. 
} 
\end{center}
\end{figure*}

In Fig.~\ref{Fig:combined}, we show the LHC and flavor bounds on $C_X^{ql}(m_b)$ for $q = (c,u)$ and $l = (e,\mu,\tau)$. 
The LHC bounds at $95\%\,\text{CL}$ for the EFT (valid for $M_\text{LQ}>10\,\text{TeV}$) and the $M_\text{LQ}=2\,\text{TeV}$ LQs are displayed in blue and cyan, respectively. 
Regarding the flavor bounds, the WC constraints from the semi-leptonic and leptonic processes are given in red and yellow, respectively. 
The WCs are taken as real unless there exists a specific direction on the complex plane of the WC, favored by the flavor bound such as $C_{V_2}^{c\tau} (m_b)$ and $C_{T}^{c\tau} (m_b)$. 
Note that the ``fit results'' (best point with $1\sigma$ uncertainty) are distinguished from the ``upper limit'' in the figure. 
Comments are written as follows. 

\begin{itemize}
 \item 
 $b \to c \ell \nu$: 
 the NP effect on this process is significant since the $|V_{cb}|$ measurement is probed from the distribution data of the process. 
 According to the previous analysis in Ref.~\cite{Iguro:2020cpg}, non-zero NP contributions of $C_{V_2, T}^{c\ell} (m_b)$ are possible at $1$--$2\sigma$ significance as shown in the figure. 
 As for the $V_2$ scenario, the present LHC bounds generally look milder than the flavor ones.
 The scalar scenarios for these light lepton modes are bounded from $\mathcal B(B_c \to \ell\nu) < 30\%$ in the same way as the tau lepton mode. 
 We can see that the LHC results for $C_T^{c\ell}$ and $C_{S_{1,2}}^{c\ell}$ are well competitive. 
 \\[-0.7em]
 \item 
 $b \to c \tau \nu$: 
 the current $R_{D^{(*)}}$ excesses are explained by the $V_1$, $V_2$, and $T$ scenarios as shown in the red bars, 
 and the LHC bounds are very comparable to these flavor-favored regions. 
 Our constraints in the EFT limit are weaker than that of Ref.~\cite{Greljo:2018tzh} but consistent with Ref.~\cite{Marzocca:2020ueu}.
 Then, it is quite significant whether the EFT limit is applicable or not to the corresponding LQ model.
 In particular, we found that the $V_2$ and $T$ solutions to the excesses are almost excluded by the LHC bounds in the EFT limits 
 whereas these scenarios are still allowed in the LQ models with $M_\text{LQ} = 2\,\text{TeV}$.
 For the scalar scenarios, see Sec.~\ref{sec:EFT}.
 \\[-0.7em]
 \item 
 $b \to u \ell \nu$: 
 At present, the flavor bound is available only from $B \to \ell\nu$. 
 For both the electron and muon modes, the scalar scenarios with $|C_{S_i}^{u\ell}| \lesssim \mathcal O(0.01)$ are allowed from the flavor bound, which is much severer than the LHC bound. 
 Regarding the vector scenarios, on the other hand, the flavor and LHC bounds are comparable for the muon mode. 
 Note that the other NP scenarios, $C_{V_{1,2},T}^{ue}$ and $C_{T}^{u\mu}$, are more constrained from our LHC study than the leptonic processes. 
 See also Ref.~\cite{Colangelo:2019axi}. 
 A comprehensive fit analysis for $B \to (\pi,\rho,a_1)\ell\bar\nu$ in the presence of the NP effects for the flavor bound would be our future work. 
 \\[-0.7em]
 \item 
 $b \to u \tau \nu$: 
 the LHC bounds are loose, naively given as $|C_{X}^{u\tau} (m_b)| \lesssim \mathcal O(1)$. 
 Nevertheless, the LHC bound is already significant for the tensor scenario, which excludes a part of the allowed regions from the flavor bound.
 We can also confirm importance of the non-EFT case as well as the other currents. 
\end{itemize}
We also present the maximum reaches at the HL-LHC in the ``no sys'' scenario of Table~\ref{Tab:LHCsummary} for comparison.
As seen, we clarified the significance of the HL-LHC sensitivity for the NP scenarios. 
For instance, the $V_2$ and $T$ solutions to the current $R_{D^{(*)}}$ excesses can be excluded.

We note that the mass of the NP model, namely LQ for the present case, is theoretically restricted apart from the above bounds. 
For instance, once we employ the unitarity bound for NP in $b \to c \tau\nu$ to explain the $R_{D^{(*)}}$ excesses, 
the NP mass is restricted as $\lesssim 9\,\text{TeV}$~\cite{DiLuzio:2017chi}.\footnote{
A loose restriction is obtained from Fig.~\ref{Fig:MassScaleCheck}.
} 
If this is the case, the EFT description is no longer appropriate for the LHC analysis with the high $p_T$ tail, but it provides the overestimated bound.
To be precise, the EFTs with $V_{1,2}$ and $S_{1,2}$ types give $> 30 \%$ $(\sim 10\%)$ severer LHC bounds than the corresponding LQ cases with $m_\text{LQ} \sim 2\, (5)\,\text{TeV}$.
As for the $T$ type, the EFT -- $2\,\text{TeV}$ LQ difference is much larger as seen in Fig.~\ref{Fig:combined}. 
Therefore, our non-EFT study is of critical importance and useful for practical NP scenarios with a NP mediator mass of $\mathcal O(1)\,\text{TeV}$. 

\begin{figure*}[t!]
\begin{center}
\includegraphics[viewport=0 0 1180 350, width=48em]{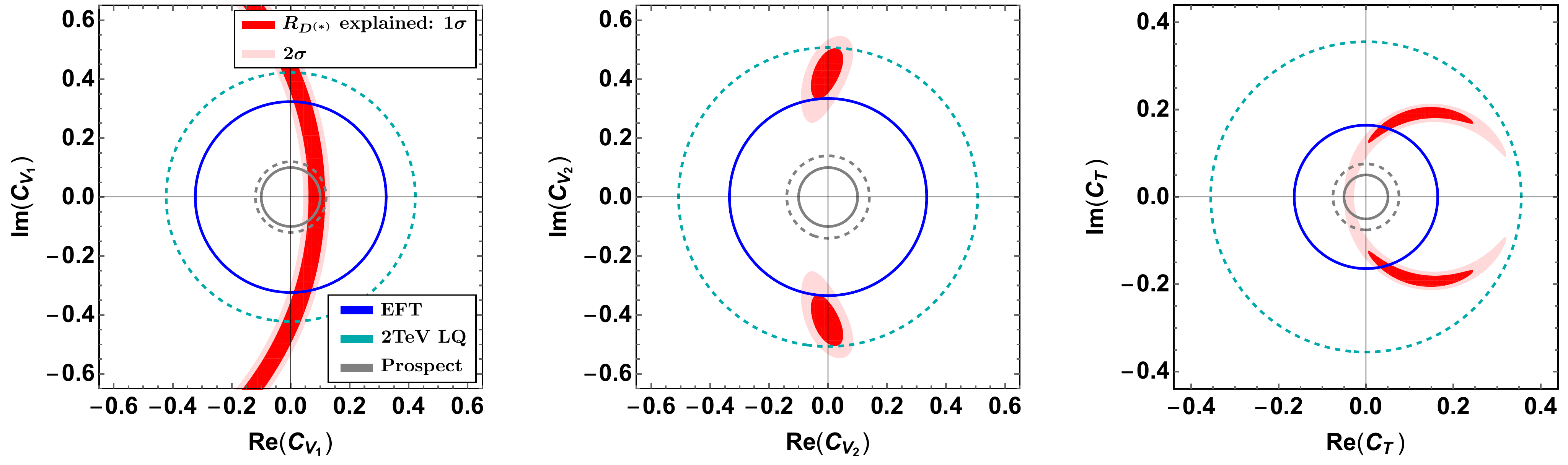} \\
\includegraphics[viewport=0 0 1180 350, width=48em]{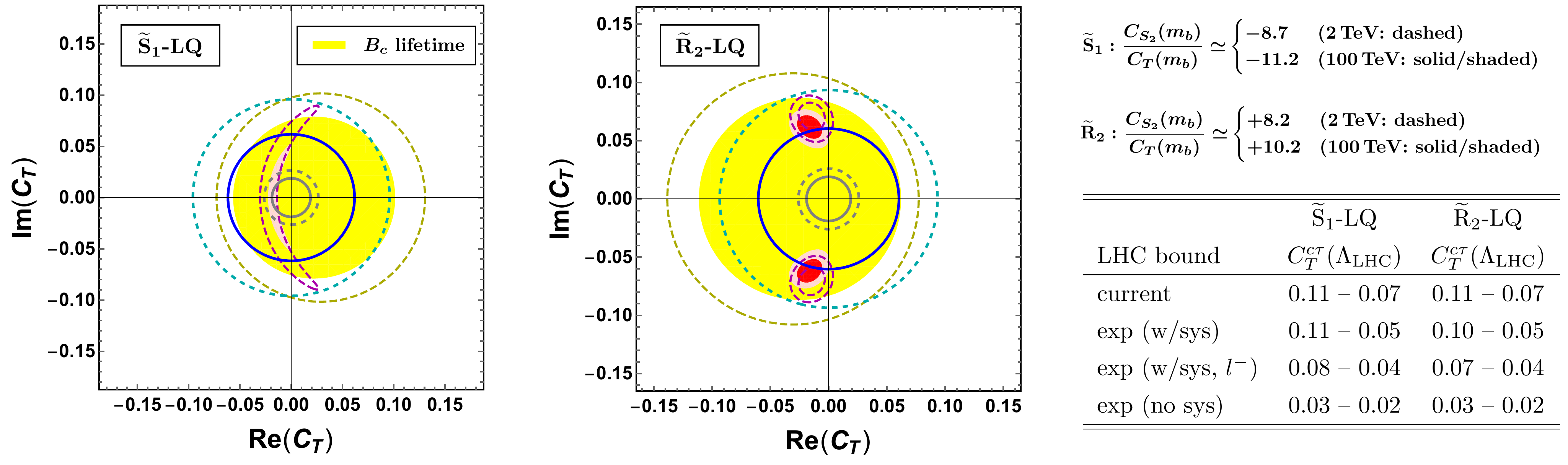} 
\caption{\label{Fig:contour}
The $R_{\Dgen}$ favored region (red:$1\sigma$\,/\,light-red:$2\sigma$) on the complex plane of $C_X^{c\tau}$ 
superimposing the LHC bounds at $95\%\,\text{CL}$ for the EFT, 2\,TeV LQ, and future prospects shown in blue, cyan, and gray, respectively. 
The left/middle/right panel is for the $V_1/V_2/T$ scenario (upper), while the left/middle for the single $\widetilde{\text{S}}_1$/$\widetilde{\text{R}}_2$ scenario and right for their LHC bounds (lower). 
} 
\end{center}
\end{figure*}

Lastly, in Fig.~\ref{Fig:contour}, we provide the favored regions on the complex plane of $C_X^{c\tau} (m_b)$ to explain the $R_{D^{(*)}}$ excesses and compare it with the current and prospect LHC bounds. 
The single $\widetilde{\text{R}}_2$ and $\widetilde{\text{S}}_1$ scenarios are also presented here since they also have the solution. 
Note that the ratio $C_{S_2}/C_T$ at the $m_b$ scale varies for the different LQ mass, which affects the allowed region.\footnote{
For the present cases, $C_{S_2} (m_b)/C_T (m_b) \simeq \{ 8.2, 10.2\}$ for $\widetilde{\text{R}}_2$ with $M_\text{LQ} = \{2, 100\}\,\text{TeV}$, while $\simeq \{ -8.7, -11.2\}$ for $\widetilde{\text{S}}_1$.
}
One finds that the $\widetilde{\text{R}}_2$ solution is almost excluded by the LHC bound for the EFT case whereas it is still viable for $M_\text{LQ} \gtrsim 2\,\text{TeV}$. 
For both cases, the HL-LHC could test this scenario. 
We can also see that the LHC bound for the $\widetilde{\text{R}}_2$ ($\widetilde{\text{S}}_1$) scenario in terms of $C_T$ is a bit severer than what is translated from that for the $V_2$ ($V_1$) operator by $C_{V_{2(1)}} \to 4C_T$ 
(unlike the $e$ and $\mu$ cases, mentioned in Sec.~\ref{sec:bound}.) 
This is due to the fact~\cite{Papaefstathiou:2011kd} that the fraction of $\tau_R$ to the visible $\tau$ decay is larger than that of $\tau_L$ at the LHC.
Thus, by determining the $\tau$ polarization we can distinguish a model that forms the $V_2$ ($V_1$) operator from $\widetilde{\text{R}}_2$ ($\widetilde{\text{S}}_1$) that generates $S_2$-$T$, although these two have (almost) the same 2 to 2 scattering kinematics. 
A similar feature can be seen in measuring the $\tau$ polarization of $\bar B \to D^{(*)}\tau\bar\nu$ at Belle~II, which could distinguish the type of LQ responsible for the $R_{D^{(*)}}$ excesses, see, {\it e.g.}, Ref.~\cite{Iguro:2018vqb}.

We can conclude that one could discover the NP signal even for a heavier LQ mass of $\sim \mathcal O(10)\,\text{TeV}$, if the $R_{D^{(*)}}$ excesses are truly caused by the NP contribution. 
Otherwise, these NP scenarios will be excluded.

\section{\boldmath Summary and Discussion}
\label{sec:conclusion}

\begin{table*}[t!]
\renewcommand{\arraystretch}{1.3}
  \begin{center}
  \scalebox{1.2}{
  \begin{tabular}{llccccc}
  \hline\hline
&  & $S_1$ 				& $S_2$ 				 & $V_1$ 					& $V_2$ 				& $T$ \\
  \hline
  $C_X^{c e} (\Lambda_\text{LHC})$ & current & $0.30$ -- $ 0.20$ & $0.33$ -- $ 0.20$ & $0.25$ -- $ 0.18$ & $0.30$ -- $ 0.18$ & $0.32$ -- $ 0.13$ \\
&exp (w/sys) & $0.35$ -- $ 0.21$ & $0.37$ -- $ 0.21$ & $0.29$ -- $ 0.19$ & $0.35$ -- $ 0.19$ & $0.37$ -- $ 0.15$ \\
&exp (w/sys, $l^-$) & $0.26$ -- $ 0.16$ & $0.27$ -- $ 0.16$ & $0.21$ -- $ 0.14$ & $0.26$ -- $ 0.14$ & $0.26$ -- $ 0.11$ \\
&exp (no sys)  & $0.12$ -- $ 0.09$ & $0.13$ -- $ 0.09$ & $0.10$ -- $ 0.08$ & $0.12$ -- $ 0.08$ & $0.11$ -- $ 0.05$ \\
  \hline
   $C_X^{c \mu} (\Lambda_\text{LHC})$ 	 &
   current & $0.41$ -- $ 0.27$ & $0.45$ -- $ 0.28$ & $0.34$ -- $ 0.25$ & $0.42$ -- $ 0.25$ & $0.35$ -- $ 0.18$ \\
&exp (w/sys)  & $0.34$ -- $ 0.22$ & $0.37$ -- $ 0.22$ & $0.28$ -- $ 0.20$ & $0.34$ -- $ 0.20$ & $0.32$ -- $ 0.14$ \\
&exp (w/sys, $l^-$)  & $0.24$ -- $ 0.16$ & $0.26$ -- $ 0.17$ & $0.20$ -- $ 0.15$ & $0.25$ -- $ 0.15$ & $0.23$ -- $ 0.10$ \\
&exp (no sys) & $0.10$ -- $ 0.08$ & $0.11$ -- $ 0.08$ & $0.09$ -- $ 0.07$ & $0.11$ -- $ 0.07$ & $0.10$ -- $ 0.05$ \\
  \hline
  $C_X^{c \tau} (\Lambda_\text{LHC})$ 	&
current   &  $ 0.45 $  --  $ 0.32 $  &  $ 0.47 $  --  $ 0.32 $  &  $ 0.42 $  --  $ 0.32 $  &  $ 0.51 $  --  $ 0.33 $  &  $ 0.42 $  --  $ 0.20 $ \\
&exp (w/sys) &  $ 0.41 $  --  $ 0.20 $  &  $ 0.43 $  --  $ 0.22 $  &  $ 0.38 $  --  $ 0.19 $  &  $ 0.48 $  --  $ 0.25 $  &  $ 0.48 $  --  $ 0.15 $ \\
&exp (w/sys, $l^-$)  &  $ 0.30 $  --  $ 0.18 $  &  $ 0.32 $  --  $ 0.18 $  &  $ 0.28 $  --  $ 0.18 $  &  $ 0.35 $  --  $ 0.19 $  &  $ 0.35 $  --  $ 0.12 $ \\
&exp (no sys)  &  $ 0.13 $  --  $ 0.10 $  &  $ 0.13 $  --  $ 0.10 $  &  $ 0.12 $  --  $ 0.10 $  &  $ 0.14 $  --  $ 0.10 $  &  $ 0.09 $  --  $ 0.06 $ \\
  \hline\hline
  $C_X^{u e} (\Lambda_\text{LHC})$ 	&
current & $0.72$ -- $ 0.37$ & $0.77$ -- $ 0.35$ & $0.59$ -- $ 0.34$ & $0.75$ -- $ 0.34$ & $0.77$ -- $ 0.23$ \\
&exp (w/sys) & $0.78$ -- $ 0.38$ & $0.84$ -- $ 0.37$ & $0.66$ -- $ 0.36$ & $0.82$ -- $ 0.35$ & $0.91$ -- $ 0.25$ \\
&exp (no sys)  & $0.33$ -- $ 0.20$ & $0.36$ -- $ 0.20$ & $0.27$ -- $ 0.19$ & $0.34$ -- $ 0.19$ & $0.29$ -- $ 0.12$ \\
  \hline
   $C_X^{u \mu} (\Lambda_\text{LHC})$ &  
current   & $0.99$ -- $ 0.58$ & $1.07$ -- $ 0.57$ & $0.83$ -- $ 0.54$ & $1.04$ -- $ 0.53$ & $0.95$ -- $ 0.35$ \\
&exp (w/sys)  & $0.81$ -- $ 0.45$ & $0.86$ -- $ 0.44$ & $0.67$ -- $ 0.42$ & $0.84$ -- $ 0.41$ & $0.83$ -- $ 0.27$ \\
&exp (no sys)  & $0.27$ -- $ 0.18$ & $0.29$ -- $ 0.18$ & $0.22$ -- $ 0.17$ & $0.28$ -- $ 0.17$ & $0.25$ -- $ 0.11$ \\
  \hline
$C_X^{u \tau} (\Lambda_\text{LHC})$ 	&
current &  $ 1.17 $  --  $ 0.65 $  &  $ 1.27 $  --  $ 0.66 $  &  $ 1.08 $  --  $ 0.72 $  &  $ 1.39 $  --  $ 0.70 $  &  $ 1.09 $  --  $ 0.41 $ \\
&exp (w/sys) &  $ 0.88 $  --  $ 0.31 $  &  $ 0.95 $  --  $ 0.30 $  &  $ 0.87 $  --  $ 0.35 $  &  $ 1.05 $  --  $ 0.34 $  &  $ 1.15 $  --  $ 0.22 $ \\
&exp (no sys) &  $ 0.36 $  --  $ 0.23 $  &  $ 0.39 $  --  $ 0.23 $  &  $ 0.33 $  --  $ 0.24 $  &  $ 0.42 $  --  $ 0.24 $  &  $ 0.29 $  --  $ 0.14 $ \\
  \hline\hline
  \end{tabular} 
 }
  \caption{   
  Summary of the current LHC bounds with $139$ fb$^{-1}$/$35.9$ fb$^{-1}$ for the $\ell$/$\tau$ modes, and the future HL-LHC potential with $3$ ab$^{-1}$, for the LQ cases of the $(2\,\text{TeV} \text{\,--\,} 100\,\text{TeV})$ masses. 
  Note that the latter case corresponds to the EFT limit.
  See the main text for the detail.
  }
  \label{Tab:LHCsummary}
  \end{center}
\end{table*}

With the help of the recent development for the $\bar B \to \Dgen$ form factors, the flavor fit analysis for the $|V_{cb}|$ determination has indicated possibilities of the non-zero NP contributions to the $b \to c \ell\nu$ current. 
On the other hand, the experimental excesses for the $R_{D^{(*)}}$ measurements have been implying an indirect NP evidence in the $b \to c \tau\nu$ current. 
A similar study regarding the $b \to u l\nu$ current also obtains upper limits on the NP effects.
These situations naturally attract us to direct searches at the LHC.

In this paper, we have considered both the Effective-Field-Theory description and the leptoquark models for all the types of the NP currents in $b \to ql\nu$ for $q = (c,u)$ and $l = (e,\mu,\tau)$, and then obtained the comprehensive flavor and LHC bounds with respect to the Wilson coefficients $C_X^{ql}$ defined as in Eq.~\eqref{Eq:effH}.

The $l^\pm + \text{missing}$ searches have been applied for this purpose, where the high $p_T$ tail of the SM background can be used to obtain the NP constraints. 
A significant point is that the EFT description is not always valid to constrain the actual NP models from the present LHC searches, 
since the NP sensitivity is gained from the transverse mass distribution at $m_T \sim 2\,\text{--}\,3\,\text{TeV}$, 
and therefore the EFT limit breaks down for the NP mass to be the same order of the $m_T$ bin. 
We have shown the clear mass dependence of the $m_T$ distribution in the LQ model for the fixed WC as in Fig.~\ref{Fig:Distributions}.

Our investigation is based on the ATLAS~\cite{Aad:2019wvl} and CMS~\cite{Sirunyan:2018lbg} analyses for $l = (e,\mu)$ and $l =\tau$, respectively. 
The LHC bounds of our results are summarized in Fig.~\ref{Fig:LHCbound} and Table~\ref{Tab:LHCsummary}. 
Then, we have seen the LQ mass dependence on the LHC bounds.
Furthermore, we have confirmed that the EFT limit is a good approximation for $M_\text{LQ} \gtrsim 10\,\text{TeV}$, 
while the vector and scalar type EFTs provide $>30\%$ ($\sim10\%$) overestimated bounds for the smaller mass of $\sim 2 (5)\,\text{TeV}$.
Regarding the tensor type, the difference is much larger such as $> 60\%$.

We have also evaluated potential of the $l^\pm + \text{missing}$ searches at the HL-LHC with $3\,\text{ab}^{-1}$, and then obtained the future projections of the HL-LHC sensitivity. 
Then we found that selecting only the $l^-$ events can potentially improve the sensitivity to $C_X^{cl}$ by $30\,\text{--}\,40\%$.
We conclude that the maximum reaches for the WC sensitivity at the HL-LHC are $|C_X^{cl}| \sim 0.1$ and $|C_X^{ul}| \sim 0.1\,\text{--}\,0.3$ mostly independent of the lepton flavor ($l=e,\mu,\tau$) and of the type of the NP operators ($X=V_1,V_2,S_1,S_2,T$).

Finally, we put the combined summary for the LHC and flavor bounds on the WCs at the low energy scale $\mu = m_b$ in Figs.~\ref{Fig:combined} and \ref{Fig:contour}. 
For some cases, one finds that the current LHC bounds are comparable with the flavor bounds. 
Here, we would like to stress again that the LHC bounds for the LQ models become milder in the case $M_\text{LQ} < 10\,\text{TeV}$ than those for the EFT, 
which is quite significant for some of the LQ scenarios when comparing it with the flavor bounds. 
In particular, the $V_2$, $T$, and $\widetilde{\text{R}}_2$-LQ (that generates the $S_2$-$T$ mixed operators) type solutions to the $R_{D^{(*)}}$ excesses are almost excluded by the LHC bounds in the EFT limits 
(which was first pointed out in Ref.~\cite{Greljo:2018tzh})
whereas they are still allowed in the LQ models with $M_\text{LQ} \gtrsim 2\,\text{TeV}$. 
This is the remarkable point of our work.

Note that the $V_1$ type NP effects for $e$ and $\mu$ can be absorbed by scaling $V_{qb}$ by $(1 + C_{V_1}^{q\ell})$ since the measurements of these processes determine the CKM elements.
Thus, it is usually hard to probe their constraints from the flavor observables. 
On the other hand, the unambiguous bound on $C_{V_1}^{q\ell}$ is obtained at the LHC, thanks to the distinct $m_T$ distribution.

We would like to propose some idea for further improvements on the $l^\pm + \text{missing}$ searches as closing remarks.  
A more dedicated analysis including an additional $b$-tagging in the $pp\to b l \nu$ mode along with the LQ mass dependence could be effective for the NP study. 
Studying multi dimensional signal distribution 
in terms of $(m_T, A_\ell, \eta)$ could enhance the NP sensitivity. 
Therefore, we would encourage both the experimental groups
to provide the $(m_T, \eta)$ distributions 
for each charge separately.


\section*{\boldmath Acknowledgement }
\label{sec:acknowledgement}
We thank  A.~Greljo, J.~Martin Camalich and J.~D.~Ruiz-Álvarez, T.~Kitahara, and and M.~Endo for valuable comments and discussion on the analysis. 
S.\,I. would like to thank the warm hospitality at KEK where he stayed during the work.
He enjoys the financial support from JSPS No.~19J10980. 
M.\,T. is supported in part by the JSPS Grant-in-Aid for Scientific Research No.~18K03611 and 19H04613. 
S.\,I. and M.\,T. are also supported by the JSPS Core-to-Core Program (Grant No.~JPJSCCA20200002).
R.\,W. is partially supported by the INFN grant ‘FLAVOR’ and the PRIN 2017L5W2PT. 

\appendix
\section{\boldmath LQ interactions and representations}
\label{sec:LQrepresentation}

Here we put a short summary for the LQ models that can contribute to $b \to ql\nu$. 
The $SU(3)_c\times SU(2)_L\times U(1)_Y$ invariant form of the LQ interaction relevant for the present process is written as 
\begin{align}
 \text{U}_1 :~~ & \bar Q_{L}^i \gamma_\mu L_{L}^j \text{U}_{1}^\mu \,,~~ \bar d_{R}^{\,i} \gamma^\mu \ell_{R}^j \text{U}_{1}^\mu \,,   \\
 \textbf{U}_3 :~~ & \bar Q_{L}^i  {\bm\sigma}^I \gamma_\mu L_{L}^j \textbf{U}_{1}^{\mu,I} \,,  \\
 \text{V}_2 :~~ & \bar d_{R}^{\,c,i} \gamma_\mu (L_{L}^j \!\cdot\! i\sigma_2 V_{2}^\mu) \,,~~ ( \bar Q_{L}^{c,i} \!\cdot\! i\sigma_2 V_{2}^\mu) \gamma_\mu \ell_{R}^j  \,, \\
 \text{S}_1 :~~ & \bar Q_{L}^{c,i} \!\cdot\! i\sigma_2 L_{L}^j \,\text{S}_1 \,,~~ \bar u_{R}^{c,i} \ell_{R}^j \,\text{S}_1 \,, \\
 \textbf{S}_3 :~~ & \bar Q_{L}^{c,i} \!\cdot\! i\sigma_2 {\bm\sigma}^I L_{L}^j \,\textbf{S}_3^I \\
 \text{R}_2 :~~ & \bar u_{R}^i (L_{L}^j \!\cdot\! i\sigma_2 \text{R}_2) \,,~~ (\bar Q_{L}^i \!\cdot\! \text{R}_2) \ell_{R}^j \,, \\
 \text{R}'_2 :~~ & \bar d_{R}^{\,i} (L_{L}^j \!\cdot\! i\sigma_2 \text{R}'_2) \,, 
\end{align}
where the quantum numbers are assigned as in Table~\ref{tab:LQ_numbers}. 
The $SU(2)_L$ doublet fields $\text{V}_2$, $\text{R}_2$, and  $\text{R}'_2$ are defined as 
\begin{align}
 \text{R}_2 = \begin{pmatrix} \text{R}_2^{5/3} \\ \text{R}_2^{2/3} \end{pmatrix} \,, 
 ~ 
 \text{R}'_2 = \begin{pmatrix} \text{R}_2^{\prime 2/3} \\ \text{R}_2^{-1/3} \end{pmatrix} \,, 
 ~
 \text{V}_2 = \begin{pmatrix} \text{V}_2^{4/3} \\ \text{V}_2^{1/3} \end{pmatrix} \,, 
\end{align} 
by indicating the electric charges for the two components. 
Given the above forms, we can extract the explicit interaction terms for $b \to ql\nu$ as shown in the main text. 
As for $\text{S}_1$ and $\text{R}_2$ that generate $C_{S_2}$ and $C_T$, we have redefined the LQ fields by taking its conjugate (just for our computation convenience). 
Note also that we have identified $\text{R}_2^{\prime 2/3}$ with $\text{R}_2^{2/3}$ in order to obtain $C_{V_2}$ of Eq.~\eqref{eq:LQV2}. 
In a proper UV theory, however, a $\text{R}_2^{\prime 2/3}$-$\text{R}_2^{2/3}$ mixing structure should be inevitable, which is beyond the scope of this work. 
\begin{table}[t]
   \begin{center}
   \begin{tabular}{cccccccc}
      \hline\hline
                        & $\text{S}_1$  & $\textbf{S}_3$  & $\text{V}_2$  & $\text{R}_2$ & $\text{R}'_2$  & $\text{U}_1$  & $\textbf{U}_3$ \\
      \hline
      spin              & 0      & 0      & 1      & 0  & 0     & 1      & 1 \\
      \hline
      $F=3B+L$          & -2     & -2     & -2     & 0  & 0    & 0      & 0 \\
      \hline
      $SU(3)_c$         & 3$^*$  & 3$^*$  & 3$^*$  & 3   & 3    & 3      & 3 \\
      \hline
      $SU(2)_L$         & 1      & 3      & 2      & 2   & 2   & 1      & 3 \\
      \hline
      $U(1)_{Y=Q-T_3}$  & 1/3    & 1/3    & 5/6    & 7/6 & 1/6    & 2/3    & 2/3 \\
      \hline\hline
   \end{tabular}
   \caption{Quantum numbers of the LQs with $SU(3)_c\times SU(2)_L\times U(1)_Y$ invariant forms.}
   \label{tab:LQ_numbers}
\end{center}
\end{table}

\section{\boldmath Angular distribution}
\label{sec:angulardistribution}

In this appendix, we discuss potential of the angular distribution for the NP search from the $l + \text{missing}$ process.

\begin{figure}[t!]
\begin{center}
\includegraphics[viewport=0 0 567 568, width=12em]{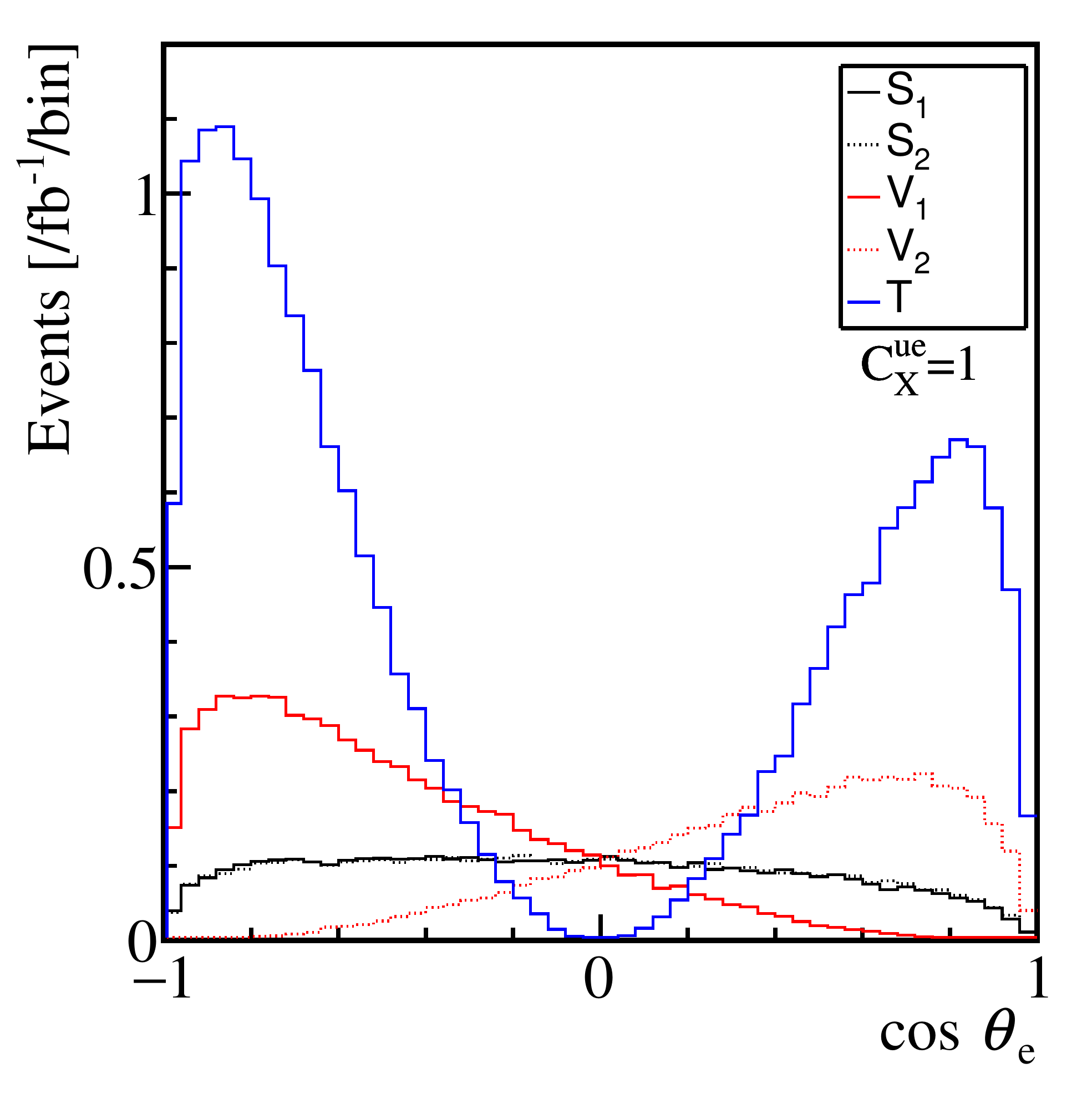} 
\includegraphics[viewport=0 0 567 568, width=12em]{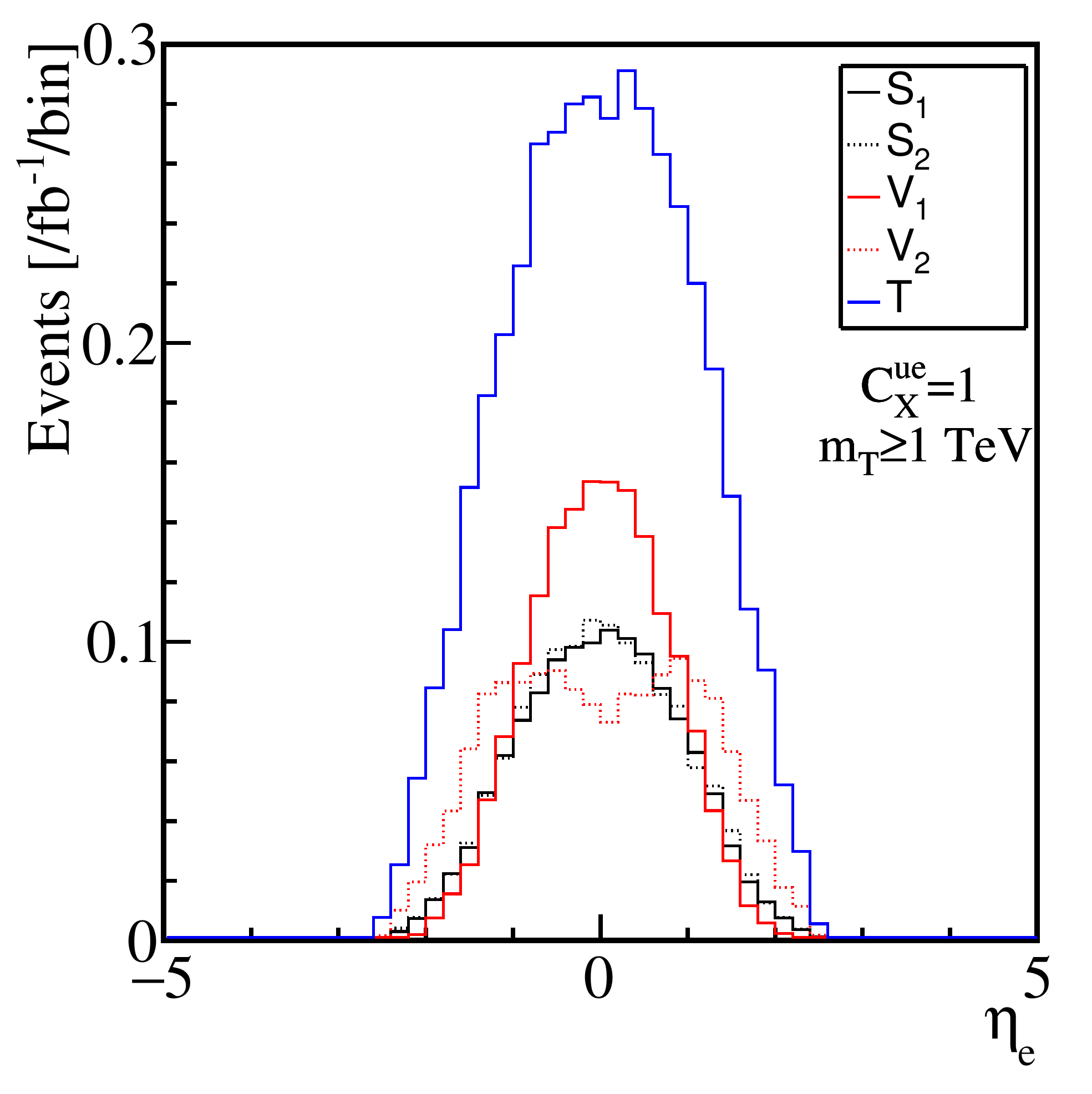} 
\caption{
\label{Fig:AngleDistribution}
Left) the $\cos\theta$ distributions for NPs in the $bu\to e\nu$ mode with $M_{NP}=100\,\text{TeV}$ ($C_X =1$).
Right) the corresponding $\eta_e$ distributions with the additional cut of $m_T > 1\,\text{TeV}$.
} 
\end{center}
\end{figure}

At first, a clear picture can be seen in the angular distribution of the observed lepton at the rest frame of the $(l \nu)$ pair as introduced in Sec.~\ref{sec:Xsl}. 
For instance, the left panel of Fig.~\ref{Fig:AngleDistribution} shows the $\cos\theta$ distribution of $pp \to e^\pm +\text{missing}$ for NP in $bu \to e\nu$, 
where $\theta$ is {\it always} defined as the angle between $e^-$ and $\bar q$ ($e^+$ and $q$) for $q = u,c$. 
Then, we see that the NP operator types are distinguishable. 
It is obvious, however, that such an angle is not directly accessible at the LHC. 
Note that the $\cos \theta$ distribution is not symmetric even for the intrinsically symmetric ones ($S_{1}$, $S_2$, $T$) since the $\eta$ cut rejects a signal event with $\cos \theta\sim 1$.

Since the angle information is partially encoded in the measurable pseudo rapidity $\eta$, we also show the $\eta$-distribution in the right panel of Fig.~\ref{Fig:AngleDistribution}.
As a result, the NP effects are degenerate, but we can still say that it is partially distinguishable.  
For instance, we can access the difference between $V_1$ and $V_2$. 
The difference stems from the fact that $u$-parton is more energetic than $\bar{b}$-parton while the $\bar u$ and $b$ partons are less energetic. 
In the $bc\to l\nu$ case, such clear differences are not observed since both ($c$, $\bar b$) and ($\bar c$, $b$) are less energetic.
Note that, for the above evaluation, all the set of the selection cuts and $m_T > 1\,\text{TeV}$ are applied.



\end{document}